\documentclass[10pt,superscriptaddress,twocolumn,amsmath,amssymb,aps,prl,showpacs,nofootinbib]{revtex4-1}
\usepackage{mathrsfs}
\usepackage{graphicx}
\usepackage{dcolumn}
\usepackage{bm}
\usepackage{amssymb}
\usepackage{amsmath}
\usepackage{paralist}
\usepackage{float}
\usepackage{mdframed}
\usepackage{booktabs}
\usepackage{mathtools}
\usepackage{graphicx}
\usepackage[colorlinks,linkcolor=blue,anchorcolor=blue,citecolor=green]{hyperref}
\usepackage{cleveref}
\usepackage{url}
\usepackage{color}
\usepackage{dsfont}

\def\be{\begin{equation}}
\def\ee{\end{equation}}
\def\bea{\begin{eqnarray}}
\def\eea{\end{eqnarray}}

 \newcommand{\braket}[1]{\langle #1 \rangle}
 
 \newcommand{\brakets}[2]{\langle #1 | #2 \rangle}
 
 \newcommand{\bra}[1]{\langle #1 |}
 
 \newcommand{\ket}[1]{| #1\rangle}

 \usepackage{ulem}

\allowdisplaybreaks[4]  
\begin{document}

\title{Microscopic origin of quantum supersonic phenomenon  in one dimension}

\author{Zhe-Hao Zhang}
\affiliation{Innovation Academy for Precision Measurement Science and Technology, Chinese Academy of Sciences, Wuhan 430071, China}
\affiliation{University of Chinese Academy of Sciences, Beijing 100049, China}

\author{Yuzhu  Jiang}
\email[]{jiangyuzhu@wipm.ac.cn}
\affiliation{Innovation Academy for Precision Measurement Science and Technology, Chinese Academy of Sciences, Wuhan 430071, China}
\affiliation{NSFC-SPTP Peng Huanwu Center for Fundamental Theory, Xi'an 710127, China}

\author{Hai-Qing Lin}
\email[]{hqlin@zju.edu.cn}
\affiliation{Institute for Advanced Study in Physics and School of Physics 
Zhejiang University, Hangzhou 310058, China}

\author{Xi-Wen Guan}
\email[]{xiwen.guan@anu.edu.au}
\affiliation{Innovation Academy for Precision Measurement Science and Technology, Chinese Academy of Sciences, Wuhan 430071, China}
\affiliation{NSFC-SPTP Peng Huanwu Center for Fundamental Theory, Xi'an 710127, China}
\affiliation{Institute for Advanced Study in Physics and School of Physics 
Zhejiang University, Hangzhou 310058, China}
\affiliation{Hefei National Laboratory, Hefei 230088,  China}
\affiliation{Department of Fundamental and Theoretical Physics, Research School of Physics,
	Australian National University, Canberra ACT 0200, Australia}


\date{\today}

\begin{abstract}

 Using the Bethe ansatz  (BA), we rigorously obtain   non-equilibrium dynamics   of  an   impurity  with a large initial momentum $Q$  in the one-dimensional (1D) interacting  bosonic medium.
 We show that  magnon and exciton-like states   obtained from the BA equations  drastically   determine  the oscillation nature of the quantum flutter  with the  periodicity given by $\tau_{\rm QF} = 2\pi/(|\varepsilon_{\rm c}(0)|- |\varepsilon_{\rm s}(0)|)$.
 Where the charge and spin dressed energies $\varepsilon_{\rm c,s}(0)$ are precisely given by the thermodynamical BA equations.
While  we  further  find  a  persistent   revival dynamics of the impurity with a larger  periodicity
$\tau_{L} = L/\left(v_{\rm c}(Q-k^*)-v_{\rm s}(k^*)\right)$ than  $\tau_{\rm QF}$, manifesting   a quantum reflection   induced by the periodic boundary conditions of a finite length $L$,
here $v_{\rm c,s}$ are  the sound velocities of charge and spin excitations, respectively, and
 $k^*$ is a characteristic momentum of the impurity to the Fermi point.
Finally, we study the application of such a magnon impurity  as a  quantum resource  for measuring  the gravitational force.
%

\end{abstract}

\maketitle



\paragraph{Introduction.}

Quantum many body systems with impurities exhibit rich collective and interference phenomena,  ranging from polaron \cite{Schirotzek:2009,Nascimbene2009,Combescot2009,Bruum2010,ZZYan2020S,Mistakidis2019PRL}, to  Bogoliubov-Cherenkov radiation \cite{Henson:2018},
shock wave \cite{Doyon2017PRL, SASimmons2020PRL, JianLi2021PRL}, Bloch oscillations \cite{Meinert:2017},    quantum flutter (QF) \cite{EDemler2012NP,EDemler2014PRL}, etc.
%
%
When an impurity is injected into a fermionic (or bosonic) medium with a speed larger than the intrinsic sound velocity, the momentum of such an impurity shows a long time oscillation behavior after a fast decay.
Such  non-equilibrium dynamical  phenomenon  was  named as ``quantum flutter'' \cite{EDemler2012NP, EDemler2014PRL,EDemler2014PRL}, showing   quasi-particle behavior of  the impurity with transitions  between the  polaron-like and exciton-like states  in the medium of free Fermi gas and  the Tonks-Girardeau (TG)Bose  gas \cite{EDemler2014PRL,BvandenBerg2016PRL2,ZZYan2020S,Mistakidis2019PRL}.
%

%
 Building on advantages of ultracold atoms, quantum simulations of  many-body phenomena   have been attracted  great deal of attention \cite{WSBakr2009N, RMPreiss2015S, IBloch2017S, CCChien2015NP,CWeitenberg2011N,JPRonzheimer2013PRL,TFukuhara2013NP,IBloch2013N,FSchmidt2018PRL,MRYang2022CM, RSChristensen2015PRL, AVashisht2022SP,XWGuan2016PRA,FMassel2013NJP,SPeotta2013PRL,NJRobinson2020JSM,JCaux2016PRL,HFroeml2019PRL,TaoShi2018PRL}.
In this scenario, one-dimensional (1D) Bethe ansatz (BA) exactly solvable models of ultracold atoms, laying  out profound many-body physics   \cite{Cazalilla:2011,Guan:2013,Guan:2022,BPozsgay2012JPA, NJRobinson2017JSM,MBZvonarev2007PRL,Guan2015CPB,TaoShi2021PRX,Andrei:1983,Lieb-Liniger,Yang:1967,Gaudin:1967,Vijayan:2020,Senaratne:2022},  provide deep insights into the phenomena  of the quasiparticles, such as  bosonic and fermionic  polarons for the  slowly moving impurities \cite{Guan-FP,McGuire:1965,XWGuan2016PRA,TLSchmidt2019PRL,ASDehkharghani2018PRL,SIMistakidis2019NJP,ZZYan2020S,Mistakidis2019PRL,JCaux2009PRA,JNFuchs2005PRL}, fractionalized magnon \cite{MBZvonarev2007PRL,Batchelor:2006} in  spin excitations,  and  supersonic impurity dynamics \cite{EDemler2012NP,EDemler2014PRL,EDemler2014PRL}  of the fast moving impurity, etc.
Such quasiparticles of polaron, magnon supersonic flutter reveal subtly different collective  features emerging  in charge, spin and spatiotemporal sectors, respectively, see reviews \cite{Amico:2021,Scazza:2022}.

On the other hand, the 1D multi-component Bose gases with a spin-independent interaction exhibit a striking feature of ferromagnetism \cite{Eisenberg:2002,Guan-Batchelor-Takahashi}.
In this regard, much effort has been devoted to experimentally manipulating  spinon and  magnon by coupling the ferromagnetic systems with an optical cavity or  external gravitational force  \cite{JNFuchs2005PRL,MBZvonarev2007PRL,Batchelor:2006,Barfknecht:2018,Patu:2018}.
However, a rigorous understanding of the  dynamics of such quasiparticles   beyond the mean field is still  challenging and highly desirable.

In this letter,  we report on  exact results   of QF  and quantum revival (QR) of  the  supersonic  impurity injected into  a medium of  1D bosonic liquid.
Building on the BA   of the 1D two-component  Bose gas  \cite{BPozsgay2012JPA, NJRobinson2017JSM}, we rigorously calculate  the time evolutions of the impurity momentum, momentum distribution and correlation function,  allowing  us to determine  exact  microscopic  states  of QF and revival.
 We show that the QF  is  caused essentially  by the coherent oscillations between the magnon and exciton-like   BA eigenstates, leading to the  periodicity given by an exact  formula (\ref{period-QF}).
It  solely depends on the interaction strength between the particles when the initial velocity of the impurity is greater than the sound velocity of the medium. 
%
 %
 %
 Whereas the finite-size energies of  magnon-like states elegantly determines  the  QR  dynamics with a larger  period  given by the analytical expression (\ref{period-L}), 
  significantly revealing    a quantum reflection of excitations with
 the sound  velocities of charge.
%
Finally we further propose a  metrological application of a magnon impurity  for measuring  the  gravitational force.

\paragraph{The model and exact solution.}
We consider the 1D two-component Bose gas described by  the Hamiltonian
\begin{eqnarray}
H
=
\int_{0}^{L}{\rm d}x
\left(
\frac{\hbar^2}{2m} \sum_{\sigma}
\partial \hat{\Psi}_\sigma^{\dagger}
\partial
\hat{\Psi}_\sigma
+
c\sum_{\sigma\sigma'}
\hat{\Psi}_{\sigma}^{\dagger}
\hat{\Psi}_{\sigma'}^{\dagger}
\hat{\Psi}_{\sigma'}
\hat{\Psi}_{\sigma}
\right),
\label{Hamiltonian}
\end{eqnarray}
 for   $N$  bosons  of the same mass $m$ with two internal spin  states $\sigma=\uparrow,\downarrow$  confined to a 1D
system of length $L$ via a $\delta$-function potential.
Where $\hat{\Psi}_{\sigma}(x)$ is the field operator of the bosons with pseudo-spin $\sigma$.
The interaction strength $c=-2/a_{\rm 1D}$ is tunable  via an effective 1D scattering length
$a_{\rm 1D}$  \cite{Olshanii_PRL_1998}.
We will use the dimensionless interaction strength $\gamma=cL/N$ and   set  $2m=\hbar =1$ in our discussions.
The model (\ref {Hamiltonian}) was solved \cite{Li-YQ:2003} by means of  the nested BA  \cite{CNYang1967PRL,Sutherland1968PRL} for arbitrary $M$ down-spins, also see  \cite{Guan-Batchelor-Takahashi}.
 Using species selective atomic systems, the related  models  were  studied experimentally on novel quantum impurity dynamics  \cite{Palzer:2009,Catani:2012,Meinert:2017}.
The eigenfunction of the model (\ref{Hamiltonian}) can be given by  the  BA wave function \cite{Li-YQ:2003} determined by  $N$ wave numbers $\{k_i\}$ with $i=1,\cdots,N$   and
$M$ spin rapidities $\{\lambda_\alpha \}$ with  $\alpha=1,\cdots,M$ satisfying the BA equations
\begin{eqnarray}
&&I_{i}
=\frac{1}{2\pi}k_{i}L-\frac{1}{2\pi}\sum_{\alpha=1}^{M}\theta(2k_i-2\lambda_\alpha)
+\frac{1}{2\pi}\sum_{j=1}^{N}\theta(k_i-k_j),
\nonumber\\
&&J_{\alpha}=\frac{1}{2\pi}\sum_{j=1}^{N}
\theta(2\lambda_\alpha-2k_j)
-\frac{1}{2\pi}\sum_{\beta=1}^{M}
\theta(\lambda_\alpha-\lambda_\beta),
\label{BAE}
\end{eqnarray}
where  $\theta(x)={2{\mathrm{atan}}(x/c)}$.
The quantum numbers are integers or half-integers,  $I_j\in \mathds{Z}+{\frac{N-M-1}{2}}$, $J_\alpha\in\{-{\frac{N-M-1}{2}},-\frac{N-M-1}{2}+1,\cdots,{\frac{N-M-1}{2}} \}$.
For a given set of quantum numbers $\{\boldsymbol{I}_N, \boldsymbol{J}_M\}$, the Eqs. (\ref{BAE}) determine the highest weight  and non-highest weight states $\ket{\boldsymbol{I}_N, \boldsymbol{J}_M,\ell} =(\hat S^-)^\ell \ket{\boldsymbol{I}_N, \boldsymbol{J}_M,0}$ with $\ell = 0$ and $\ell = 1,\cdots,N-2M$, respectively, see Supplemental  material (SM) \cite{SM}.
The  energy and momentum of the model are given by
\begin{equation}
 E=\sum_i {k_i}^2,\qquad  K = \frac{2\pi}{L} \Big(\sum_iI_i-\sum_\alpha J_\alpha\Big),
 \label{EandK}
\end{equation}
respectively.
%
%

%

%
%
%


\begin{figure}[t]
	\centering
	\begin{center}
		\includegraphics[width=1.\linewidth]{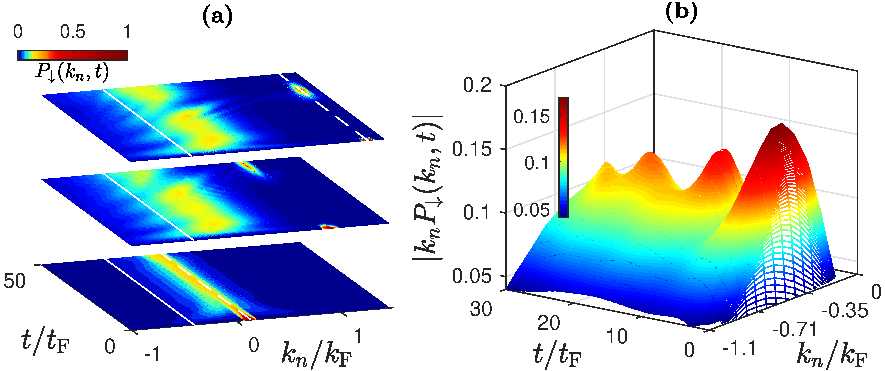}
	\end{center}
	\caption
	{(a) The  time evolution of the impurity momentum distribution $P_\downarrow$ for different values of  $Q/k_{\rm F}=1.33$, $1$, and $0.07$ from top to bottom.
	 For $Q\geq k_{\rm F}$,
	 several persistent peaks occur (white dashed line), showing   long time QR at $k_n\approx Q$.
    When $k_n\approx - 0.4  k_{\rm F}$, the undulant behaviour (white solid line)  manifests  the nature of QF.
	(b)  shows the time evolution of impurity momenta at different $k_n$s, namely
	$k_nP_{\downarrow}(k_n,t)$.
 The undulant peaks near $k_n\approx -0.4 k_{\rm F}$ shows  large  oscillations of the revival.
	In both figures we set  $\gamma=10$ and $N=30$.
	}
	\label{f:Pdown}
\end{figure}

\paragraph{Initial state, density matrix and form factor.}
We consider the ground state of $N-1$ spin-up delta-function interacting bosons $\ket{\varOmega}$ as a medium,  and one spin-down atom with a wave function $\phi_\downarrow(x)$ as an injected impurity.
%
This gives an  initial state $\ket{\varPhi_{\rm I}} = \int {\rm d} x  \phi_{\downarrow}(x) \hat{\Psi}_{\downarrow}^{\dagger}(x)
\ket{\varOmega}$.
Thus the time evolution of density matrix of the spin-down boson is given by
\begin{eqnarray}
 \rho_{\downarrow}(x,x',t)
&=&
\frac
{\bra{\varPhi_{\rm I}}
\hat{\Psi}_{\downarrow}^{\dagger}(x,t)
\hat{\Psi}_{\downarrow}^{}(x',t)
\ket{\varPhi_{\rm I}}
}
{\langle\Phi_{\rm I}\rangle}
\nonumber\\
%
&=&
\sum_{\alpha,\alpha'}
{\rm e}^{ {\rm i}(E_{\alpha}-E_{\alpha'})t}
A^*_{\alpha} A_{\alpha'} \rho_{\downarrow}^{\alpha\alpha'}(x,x'),
\label{rho}
\end{eqnarray}
where $\hat{\Psi}_{\sigma}^{\dagger}(x,t)={\rm e}^{{\rm i}\hat{H}t}\hat{\Psi}_{\sigma}^{\dagger}(x){\rm e}^{-{\rm i}\hat{H}t}$,
$\rho_{\downarrow}^{\alpha\alpha'}(x,x')= \bra{\alpha}$ $\hat{\Psi}_{\downarrow}^{\dagger}(x)  \hat{\Psi}_{\downarrow}^{}(x') \ket{\alpha'} /{\sqrt{\langle \alpha\rangle \langle \alpha' \rangle}}$
is the matrix element of the density operator and $A$ is the overlap between the initial state and the eigenstates, $A_{\alpha }
=\braket{\alpha| \varPhi_{\rm I}}/\sqrt{\braket{\alpha}\braket{\varPhi_{\rm I}}}$.
$\ket{\alpha}$ is the highest weight state denoted by  $\ket{\boldsymbol{I}_N,J,0}$ or the non-highest weight state $\ket{\boldsymbol{I}_N, \boldsymbol{J}_0,1}$,  $E_{\alpha}$ is the energy of the state $\ket{\alpha}$, here $\boldsymbol{J}_0$ denotes  an empty set.
The non-highest state  was largely ignored in literature, here we notice its nontrivial contributions to the impurity dynamics, see SM \cite{SM}.
Using the determinant representation   \cite{BPozsgay2012JPA,Caux:2006,Caux:2007,Caux:2009,Song:2022-1,Li:2023,BPozsgay2012JPA}, we  precisely calculate the  overlapping integral $A_\alpha$ and the density matrix $\rho_{\downarrow}^{\alpha\alpha'}(x,x')$.
Without losing generality, we will take  the impurity wave packet  as  a plane wave with momentum $Q$, i.e. $\phi_{\downarrow}(x)={\rm e}^{{\rm i}Qx}$, and the  corresponding  the projected states have a fixed total momentum, i.e., $K=Q$,
the momentum is conserved in the  states $\ket{\alpha}$.
This naturally gives a selection rule of the overlap $A_\alpha$ and the matrix element.
%
%
With the help of   the sum rule of $A_\alpha$,  we may  select enough essential  states such that the sum rule is very close to 1,  see SM \cite{SM} for details.
\begin{figure*}[th]
	\centering
\begin{center}
 \includegraphics[width=\linewidth]{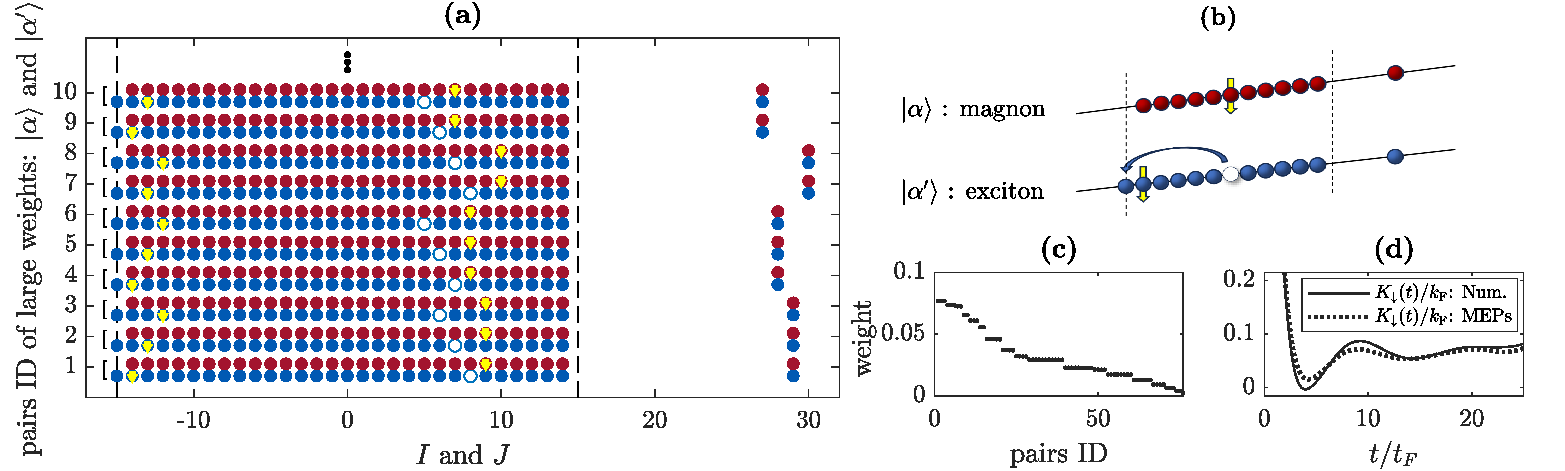}
\end{center}
\caption{
    Selected microscopic states  of the QF with  $Q=1.33k_F$, $N=30$, and $\gamma=10$.
	(a) The macroscopic MEPs of  $\ket{\alpha'}$ and  $\ket{\alpha}$  have  large contributions to the QF.
	The dots always stand for the quantum numbers $\boldsymbol{I}_N$  and the yellow-arrow  $\downarrow$ indicates  the position of   the quantum number $J$.
	The rows with red dots  denote the magnon-like states and  the ones with blue dots denote the  exciton-like states.
	(b)  Schematic illustration of the magnon- and exciton-like states.
	The former  (red)  has an emitted particle outside the Fermi sea, and  a spin-down particle  sits near the  center.
	In the latter  (blue),  a particle in the deep  Fermi sea excites onto the Fermi surface, while  a down-spin sits near the left Fermi point  and the emitted particle sits outside the Fermi sea.
	(c) We show the contributions from the MEPs pairs with  high weights $|A_\alpha|^2+|A_{\alpha'}|^2$. 
	Here we normalized the sum rule weights by the largest one.
	(d) The black solid line stands for the numerical  result  of $K_\downarrow(t)$, whereas the black dotted line shows  the  result obtained from  the selected MEPs.
	%
}
\label{f:QF-states}
\end{figure*}
Here we would like to emphasize that, to analyze the microscopic origin  of physical phenomena,  a subspace with proper truncation can be efficient 
 \mbox{\cite{Burovski:2014,Gamayun:2018,Gamayun:2020,Caux:2020,Gamayun:2023}}.
The single particle-hole excitations with hole nearby the Fermi surface present the most essential ones for the  long time limit behavior \cite{Gamayun:2018}.
In contrast, here we selected  the projected pairs  of states, i.e. the magnon- and exciton-like states, to discuss the periodicities of QF and QR.

\begin{figure}[t]
 \begin{center}
  \includegraphics[width=\linewidth]{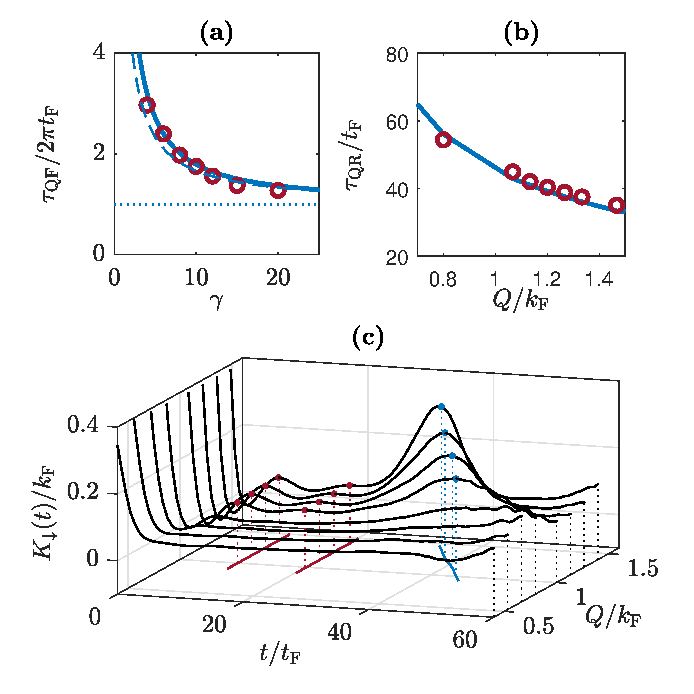}
 \end{center}
\caption{
(a) The blue solid  line shows  the period of the QF calculated by using Eq.  (\ref{period-QF}),
showing a good agreement with the numerical result (circles) obtained  from  the states of MPs pairs.
 The  blue dotted (long  dashed)  line denotes the result of $\tau_{\rm QF}=2\pi t_{\rm F}$  in the Tonks  limit (strong coupling region $\tau_{\rm QF} = 2\pi t_{\rm F}(1+20/3\gamma)$).
 (b) The larger periodic  revival of  $K_\downarrow(t)$ (blue solid) obtained  from Eq. (\ref{period-L}) agrees well with the numerical result (circles) obtained  from  the states of MPs pairs.
 (c) shows the  dynamics of  the QF  and  QR of the impurity with different values of  initial momentum $Q$  for  fixed   interaction strength $\gamma=10$ and particle number $N=30$.
The red and blue   solid lines show the periodicities of the QF and QR, respectively.}
 %
\label{f:tauQF-K-t}
\end{figure}

\paragraph{Microscopic origin of Quantum flutter. }
The time evolution of the impurity momentum is given by
$K_{\downarrow}(t) = \sum_{k_n} k_n P_{\downarrow}(k_n,t)$, here
the Fourier component
$k_n=2n\pi/L$, $n=0,\pm1,\cdots$  and the probability of impurity in  momentum space is given by
\begin{eqnarray}
P_{\downarrow}(k_n,t)
&
=&
\int \rho_{\downarrow}(x,0,t){\rm e}^{{\rm i}k_nx} {\rm d} x \label{P-down}\\
&
=&
\sum_{\alpha,\alpha'}
{\rm e}^{ {\rm i}(E_{\alpha}-E_{\alpha'})t}
A^*_{\alpha} A_{\alpha'}\int \rho_{\downarrow}^{\alpha\alpha'}(x,0){\rm e}^{{\rm i}k_nx} {\rm d} x.\nonumber
\end{eqnarray}
Using the BA solution of Eq. (\ref{BAE}) and its form factor, see \cite{SM}, we rigorously calculate the time evolution of the distribution $P_\downarrow(k_n,t)$ in FIG \ref{f:Pdown} (a), showing
a QF wave-like oscillation  near $k_n=-0.4 k_F$ and the  revival at the original momentum $k_n=Q$.
In  FIG. \ref{f:Pdown}  (b), we shows that  a coherent oscillation of  the impurity momentum $K_\downarrow(t)$   occurs soon after a quick decay, also see later discussion in FIG. \ref{f:QF-states} (d) and FIG. \ref{f:tauQF-K-t} (c).
The QF  dynamics drastically  comes from the coherent transition between the magnon excitations and  particle-hole collective excitations resulted in  from the  impurity scattering  with the atoms in the  interacting medium, which we simply call magnon- and exciton-like  states, respectively.
%
%

%
 From  Eq. (\ref{P-down}),  we observe that  the time evolution of the momentum distribution  $P_{\downarrow}(k_n,t)$ depends on  the energy differences $E_{\alpha}-E_{\alpha'}$ between  the particle-hole excitations and exciton-like excitations, which determine the  the oscillation periodicity  $\tau_{\rm QF}\sim2\pi/|E_\alpha-E_{\alpha'}|$.
This  naturally suggests  a mechanism for the supersonic behaviour, i.e.  coherent transition between the  states  $\{\ket{\alpha},\ket{\alpha'}\}$.
%
%
%
%
%
{Being guided by the sum rule weights,   in FIG. \ref{f:QF-states} (a),
for $N=30$, we find that the $10$ pairs of states $\{\ket{\alpha},\ket{\alpha'}\}$ with the high sum rule weights mainly capture the dynamics of  $P_\downarrow(k_n,t)$ and $K_\downarrow(t)$.}
Here we used  the same setting as that for the  FIG. \ref{f:Pdown}.
While FIG. \ref{f:QF-states} (b) presents  a schematic illustration of the magnon- and exciton-like states of  these pairs (states $\ket{\alpha}$  and $\ket{\alpha'}$ ).
 %
%
In FIG. \ref{f:QF-states} (c), we  give  the high weights of these excitation pairs $|A_\alpha|^2+|A_{\alpha'}|^2$,
showing   the  contribution of each  magnon-exciton  pair (MEP)  to the dynamical evolution of the impurity.
In FIG. \ref{f:QF-states} (d), we show that the $K_\downarrow (t)$
obtained from the selected MEPs (black dotted line) coincides  with the numerical result (black solid line) from the BA wave function.
 This remarkably indicates  that the microscopic MEPs   result  in the  coherent  transitions between $\ket{\alpha}$ and $\ket{\alpha'}$ states with  the almost  same energy difference  \cite{SM}.

We further obtain   the periodicity  of the QF
\begin{eqnarray}
\tau_{\rm QF} = \frac{2\pi}{|\varepsilon_{\rm c}(0)|-|\varepsilon_{\rm s}(0)|}.
\label{period-QF}
\end{eqnarray}
Where  the charge  and spin dressed energies are determined  by  $\varepsilon_{\rm c}(k)= k^2 - \mu - \int_{-k_0}^{k_0} a_{2}(k-k')\varepsilon_{\rm c}(k'){\rm d}k'$, $\varepsilon_{\rm s}(\lambda)= - \int_{-k_0}^{k_0} a_{1}(k-\lambda)\varepsilon_{\rm c}(k){\rm d}k$, respectively \cite{SM}.
Here we denoted  $a_n(x)=\frac{1}{2\pi}\frac{nc}{(nc/2)^2+x^2}$, $\mu$ is the chemical potential, $k_0$ is the Fermi point (cut-off) of charge quasimomentum $k$, $\varepsilon_{\rm c}(k_0)=0$ and $\varepsilon_{\rm s}(\pm\infty)=0$.
 FIG \ref{f:tauQF-K-t} (a) shows   the oscillation  period v.s.  the  interaction strength,
confirming an agreement between the analytical result (blue solid line) Eq. (\ref{period-QF})  and the numerical  result (circles).
The period of the QF   decreases  with  an increase of the interaction $\gamma$.
For a strong coupling, we have $\tau_{\rm QF} = 2\pi t_{\rm F}(1+20/3\gamma)$ (long dashed line), where the $t_{\rm F}=1/E_{\rm F}$ with the Fermi energy $E_{\rm F}=k_{\rm F}^2$.
In the Tonks limit, i.e. $\gamma \to \infty$, $\Delta E_{\rm QF}= E_{\rm F}$ and thus $\tau_{\rm QF}= 2\pi t_{\rm F}$ (blue dotted line).
%

Moreover, we note that the  period of the QF  dose not depend on the injected momentum $Q$.
We calculate the QF dynamics  for several values of the injected momenta  and find that the QF  always appears for $Q>k_F$ and oscillating  amplitude rises  slightly  as the increase of $Q$, see FIG \ref{f:tauQF-K-t} (c).
In addition, using the Gaussian impurity wave packet \cite{SM}, we further observe that  the motion of  mass center of the impurity $ X_\downarrow(t)= \bra{\varPhi_{\rm I}} \hat x(t) \ket{\varPhi_{\rm I}}/\braket{\varPhi_{\rm I}}$ coincide with the evolution of the impurity momentum, namely,
$ \frac12\partial_t X_\downarrow(t) = K_\downarrow(t)$, see \cite{SM}.
\begin{figure}[t]
 \begin{center}
  \includegraphics[width=\linewidth]{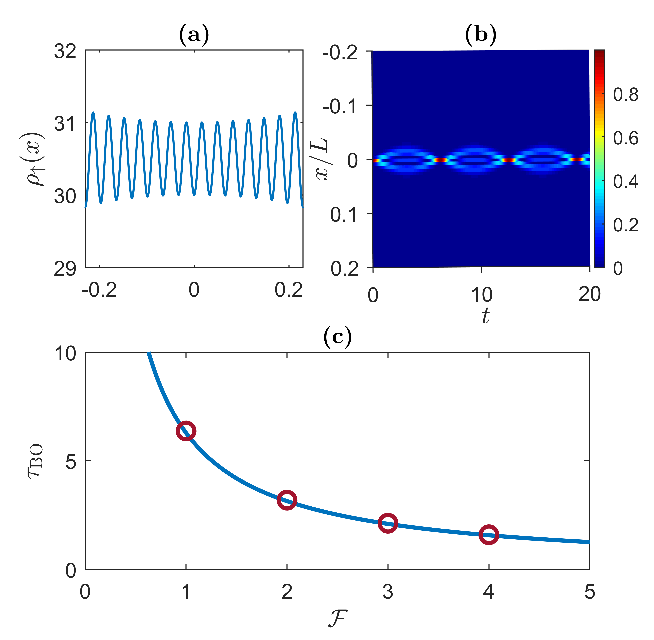}
\end{center}
\caption{
The Bloch oscillations  of the magnon impurity. We set $N=L=30$, and $c_{\rm IB}=1$ for our numerical calculation.
(a) shows the density profile $\rho_{\uparrow}(x)$ of the medium in  the ground state of the Lieb-Liniger Hamiltonian $H_{\rm LL}$ with  an infinite interaction  strength.
(b) shows the time evolution of density distribution of the impurity under the Hamiltonian (\ref{H_Im}) with $\mathcal{F}=1$.
(c) The solid line shows  the period of BO $\tau_{\rm BO}={2\pi}/{\mathcal{F}}$ with respect to the gradient field ${\cal F}$, the circles denotes the  numerical results.
}
\label{f:quantum-walk}
\end{figure}

\paragraph{Quantum revival and Bloch oscillation }

		%

FIG. \ref{f:Pdown} (a) showed  another periodic revival  behavior appearing  along the line $k_n=Q$,  also see the evolution of momentum $K_\downarrow(t)$  in FIG. \ref{f:tauQF-K-t} (c).
This striking feature is essentially related to  the quantum reflection of excitations  induced by the periodic boundary conditions.
Using BA equations  (\ref{BAE}), we determine a set of pairs of  magnon-like states with the same minimum momentum  difference that have large sum rule weights for the QR dynamics of the impurity, see \cite{SM}.
%
%
Here we precisely determine   that  pairs of magnon states  have a minimum momentum difference $\Delta p = 2\pi/L$ lead to an energy difference $\Delta E_L =[v_{\rm c}(Q-k^*)-v_{\rm s}(k^*)] \Delta p$, where  the sound  velocities of the charge  and magnon excitations are given by   $v_{\rm c,s}(p)=\partial  E_{\rm c,s}(p)/\partial p$,  and    $E_{\rm c,s}(p)$  the single particle  dispersions of charge and spin, respectively.
Consequently,  the period of QR is given by
\begin{eqnarray}
\label{period-L}
\tau_{\rm QR}
=\frac{L}{v_{\rm c}(Q-k^*)-v_{\rm s}(k^*)},
\end{eqnarray}
when $Q>k_{\rm F}$, the $k^*$ can be numerically determined by the magnon-like state with the largest weight $|A_{\alpha}|^2$,
namely, we have $k^*=(1-2J/N)k_{\rm F}$, here $0<k^*<k_{\rm F}$.
The value  $k^*$ is related to the quantum number of the BA state with the highest projection weight.
 A deep insight into the  QR  can be conceived from the impurity momentum  $\tilde K_\downarrow(E)$ in frequency space,  see SM \cite{SM}.
The QR is also observed in  the single particle  propagator \cite{SM}.
%
 %

 On the other hand, the model (\ref{Hamiltonian})  also provides a promising  metrological resource  for measuring gravitational force via the following experimentally realizable Hamiltonian \cite{Meinert:2017}
$
H_{\rm BO}
=
H_{\rm LL}
+
T_{\rm Im}
+
\int_{0}^{L}{\rm d}x
\left(
c_{\rm IB}\sum_{\sigma=\uparrow}
\hat{\Psi}_{\sigma}^{\dagger}
\hat{\Psi}_{\sigma}
\hat{\Psi}_{\downarrow}^{ \dagger}
\hat{\Psi}_{\downarrow}
+\mathcal{F}x \hat{\Psi}_{\downarrow}^{ \dagger}
\hat{\Psi}_{\downarrow}
\right){\rm d}x
$,
where $H_{\rm LL}=\int_{0}^{L}{\rm d}x\big(\partial\hat{\Psi}_{\uparrow}^{\dagger}\partial\hat{\Psi}_{\uparrow}+c\hat{\Psi}_{\uparrow}^{\dagger}\hat{\Psi}_{\uparrow}\hat{\Psi}_{\uparrow}^{\dagger}\hat{\Psi}_{\uparrow}\big)$ is the spinless  Lieb-Liniger gas of model  (\ref{Hamiltonian}), $T_{\rm Im}$ is the kinetic energy of the impurity, $c_{\rm IB}$ denotes  the interaction between the impurity and the  medium, and $\mathcal{F}$ is the gravitational potential.
Here we  consider  the TG gas as the interacting medium $H_{\rm LL}$ such that  the density distribution $\rho_{\uparrow}(x)=\langle\hat{\Psi}^{\dagger}_{\uparrow}(x)\hat{\Psi}^{}_{\uparrow}(x)\rangle$ in the medium naturally provides a periodic potential as a 1D  lattice.
Using the result of the wave function given  in \cite{XWGuan2016NJP},   we demonstrate such an existence of an ideal 1D lattice structure of the density profile in the central region  of the medium, see  FIG. \ref{f:quantum-walk} (a).
Based on  this observation,  we can  further  regard  the effective Hamiltonian $H_{\rm BO}$ as
\begin{eqnarray}
	H_{\rm BO}
	\approx
	T_{\rm Im}
	+
	\int_{0}^{L}\bigg(c_{\rm IB}\hat{\rho}_{\downarrow}(x)\rho_{\uparrow}(x)+\mathcal{F}x\hat{\rho}_{\downarrow}(x)\bigg){\rm d}x.
	\label{H_Im}
\end{eqnarray}

In  FIG. \ref{f:quantum-walk} (b), we set the initial state with the  impurity  at the center of the center of the medium$(x=0)$.
The impurity  is accelerated by gravitational force  $\mathcal{F}$, and the impurity  interaction  with the medium $c_{\rm IB}\rho_{\uparrow}(x)$ provides a periodic potential, keeping the momentum of the impurity in the Brillouin zone.
Consequently, we observe the Bloch oscillation (BO) with the  periodicity  given by $\tau_{\rm BO}=2\pi/\mathcal{F}$, see  Fig.~\ref{f:quantum-walk} (c).
This provides a useable metrological application of the magnon impurity for  measuring  the gravitational constant $g$ through the BO.
From a perspective of the quantum resource theory,  applications of magnon  impurities  in sensoring gravitational force  and external magnetic field are plausible
\cite{Wan:2024,Puhan:2017,cs:XWGuan2021PRL}.

In summary, using BA and form factor,  we have rigorously determined microscopic states for QF and QR of a supersonic  impurity  injected into  the 1D medium of interacting bosons.
 We have obtained explicit expressions of  the  periods of the QF  (\ref{period-QF}) and  the QR (\ref{period-L}),  revealing deep insights into the   coherent features of  the magnon- and exciton-like states  in the course of impurity scattering with the interacting medium.
Based on the current experimental capability of realizing the   1D impurity problems \cite{Palzer:2009,Catani:2012,Meinert:2017}, measurement of the supersonic behaviour of the model (\ref{Hamiltonian}) can be readily  implemented  through highly elongated 1D systems of selective ultracold atoms.
Finally,  we   have proposed a metrological  application of  the quantum impurity  in sensoring  the gravitational force.
Our results provide  an extended  understanding   of  the quantum supersonic impurities   in the  1D interacting medium of Luttinger liquids.
%
%
%

\section*{Acknowledgement}
  X.W.G and Y.Z.J. are supported by the NSFC key grants No. 12134015, No. 92365202,
 No. 12121004, No. 12175290 and  the National Key R\&D Program of China under grants No. 2022YFA1404102.
 They also partially supported by the Innovation Program for Quantum Science and Technology 2021ZD0302000,  the Peng Huanwu Center for Fundamental Theory, No. 12247103, and the Natural Science Foundation of Hubei Province 2021CFA027.  H.Q.L  acknowledges financial support from NSFC12088101.
 %


\clearpage\newpage

\setcounter{equation}{0}
\setcounter{figure}{0}
\setcounter{table}{0}
\renewcommand{\theequation}{s\arabic{equation}}
\renewcommand{\thefigure}{s\arabic{figure}}
\renewcommand{\thetable}{s\arabic{table}}

\begin{titlepage}
\begin{center}
{\large \bf
Microscopic origin of quantum supersonic phenomenon  in one dimension\\~\\
--- Supplementary materials}\\~\\
{Zhe-Hao Zhang, Yuzhu Jiang, Hai-Qing Lin, and Xi-Wen Guan}\\~
\end{center}

\end{titlepage}

\section{S1. The one-dimensional  two-component Bose gas}

The model Eq. (1) in the main text describes the one-dimensional  (1D)  two-component Bose gases with a delta-function interaction.
As a solvable  many-body problem, its Hamiltonian reads
\begin{equation}
 \hat H =- \sum_{i=1 }^{N}\frac{\partial^{2}}{\partial x_i^{2}}+2c\sum_{i<j}\delta(x_i-x_j),
 \label{es:H}
\end{equation}
where $N$ is the total particle number, $c$ is the interaction strength and $L$ is  length of the system.
Here we take the periodic boundary conditions and the total momentum $\hat K$ is conserved.
The eigenstate of $N$ particles with $M$ spin-down bosons of the  model Eq. (1) in the main text is given by
\begin{eqnarray}
 \ket{\varPsi} =
 &&\int {\rm d}\boldsymbol{x} \varPsi(\boldsymbol{x})
 \hat\Psi_\downarrow ^\dag(x_1) \hat\Psi_\downarrow ^\dag(x_2) \cdots \hat\Psi_\downarrow ^\dag(x_M)
 \nonumber\\
 &&\times \hat\Psi_\uparrow ^\dag(x_{M+1}) \hat\Psi_\uparrow ^\dag(x_{M+2})\cdots \hat\Psi_\uparrow ^\dag(x_N) \ket{0},
\end{eqnarray}
where $\hat \Psi_{\uparrow,\downarrow}^\dag(x)$ are the field operators of spin-up and spin-down bosons, respectively, $\varPsi(\boldsymbol{x})$ denotes  the Bethe ansatz  (BA) wave function of the first quantized Hamiltonian (\ref{es:H}). Here  we denoted $\boldsymbol{x}=\{x_1,x_2,\cdots,x_N\}$,  $\int {\rm d}\boldsymbol{x}=\int_0^L {\rm d}x_1 \int_0^L {\rm d}x_2 \cdots \int_0^L {\rm d}x_N$ and $\ket{0}$  stands for the vacuum state.

This model was exactly solved by the BA  \cite{cs:CNYang1967PRL,Sutherland1968PRL,cs:SJGu2002IJMPB} and the Bethe ansatz equations (BAE) are given by
\begin{equation}
\label{es:BAE-origin}
\begin{split}
 &{\rm e}^{{\rm i} k_jL} = -
 \prod_{j'=1}^N \frac{k_j-k_{j'}+{\rm i} c}{k_j-k_{j'}-{\rm i} c}
 \prod_{\alpha = 1}^M \frac{k_j-\lambda_{\alpha}-{\rm i}c /2}{k_j-\lambda_{\alpha}+{\rm i}c/2},
 \\
 &\prod_{j=1}^N
 \frac{\lambda_\alpha-k_{j}-{\rm i}c/2}{\lambda_\alpha-k_{j}+{\rm i}c/2}
 =-\prod_{\beta=1}^M
 \frac{\lambda_\alpha-\lambda_{\beta}-{\rm i}c}{\lambda_\alpha-\lambda_{\beta}+{\rm i}c},
 \end{split}
\end{equation}
where $k_j$ is the wave number, $\lambda_\alpha$ is the spin rapidity, $j=1,2\cdots,N$ and $\alpha=1,2\cdots,M$.
 Eqs. (2) in the main text were obtained from  the logarithm form of  the BAE (\ref{es:BAE-origin}).
 Both the energy and momentum are conserved and  they are given by
\begin{equation}
 E =\sum_{j=1}^N k^2_j,~~~ K =\sum_{j=1}^N k_j,
\end{equation}
respectively.
Moreover, the total momentum can  be calculated by the quantum numbers of the logarithm form of BAE, see Eq. (3) in the main text.

\section{S2. Quantum dynamics of the supersonic impurity}

We first  discuss the evolution of impurity momentum  injected into a bosonic quantum medium.
The medium is the ground state of of $N-1$ spin-up bosons $\ket{\varOmega}$, and the impurity is a spin-down particle with a wave function $\phi_{\downarrow}(x)$.
We define the initial state of the supersonic impurity
\begin{equation}
 \ket{\varPhi_{\rm I}} = \int_0^L {\rm d} x \phi_{\downarrow}(x) \hat{\Psi}_{\downarrow}^{\dagger}(x)
 \ket{\varOmega}.
 \label{es:initial-state}
\end{equation}
The time evolution of the impurity momentum is defined by
\begin{equation}
K_{\downarrow}(t)
= \sum_{k_n} k_n P_\downarrow(k_n,t),
\end{equation}
where $P_\downarrow(k_n,t)$ is the momentum distribution
\begin{eqnarray}
 P_\downarrow(k_n,t)
 && =
 \frac{1}{L}\int_0^L {\rm d}x \int_0^L  {\rm d}x' {\rm e}^{-{\rm i}k_n (x-x')}
 \nonumber\\
 && \times
 \frac
{\bra{\varPhi_{\rm I}}
\hat{\Psi}_{\downarrow}^{\dagger}(x,t)
\hat{\Psi}_{\downarrow}^{}(x',t)
\ket{\varPhi_{\rm I}}
}
{\langle\varPhi_{\rm I}\rangle},
\end{eqnarray}
$\hat{\Psi}_{\downarrow}^{\dagger}(x,t) =  {\rm e}^{{\rm i}(\hat{H}t - \hat Kx)}\hat{\Psi}_{\downarrow}^{\dagger}(0){\rm e}^{-{\rm i}(\hat{H}t - \hat Kx)}$ and $k_n = 2n\pi/L$, $n=0,\pm1,\cdots$.
Insert three complete sets of eigenstates into $P_\downarrow(k_n,t)$, we get
\begin{eqnarray}
 P_\downarrow&&(k_n,t)
  =
 k_n L \sum_{\alpha\alpha'\beta}{\rm e}^{{\rm i}(E_{\alpha} - E_{\alpha'})t}
 \delta_{k_n,K_{\alpha}-K_{\beta}}\delta_{K_{\alpha},K_{\alpha'}}
\nonumber\\
 &&\hspace{15pt} \times
 \frac
{\brakets{\varPhi_{\rm I}}{\alpha}
\bra{\alpha}
\hat{\Psi}_{\downarrow}^{\dagger}(0)\ket{\beta}\bra{\beta}
\hat{\Psi}_{\downarrow}^{}(0)
\ket{\alpha'}
\brakets{\alpha'}{\varPhi_{\rm I}}
}
{\braket{\varPhi_{\rm I}} \braket{\alpha}\braket{\beta} \braket{\alpha'}},
\end{eqnarray}
where $\ket{\alpha}$, $\ket{\alpha'}$ and $\ket{\beta}$ are eigenstates of the Hamiltonian and momentum, namely, $H\ket{\alpha}=E_\alpha\ket{\alpha}$ and $\hat K \ket{\alpha}=K_\alpha\ket{\alpha}$.
Thus the time evolution of impurity momentum can be written as
\begin{equation}
\begin{split}
 K_{\downarrow}(t)
 & = L \sum_{\alpha\alpha'}
 {\rm e}^{{\rm i}(E_{\alpha} - E_{\alpha'})t} K_{\alpha\alpha'},
 \\
 K_{\alpha\alpha'}
 & = \sum_{\beta} (K_{\alpha}-K_{\beta})
 A^*_\alpha B^*_{\alpha\beta}B_{\alpha'\beta}A_{\alpha'}\delta_{K_\alpha,K_{\alpha'}},
\end{split}
\label{es:kd-1}
\end{equation}
where $A$ is the overlap between the initial state and the eigenstate, $A_{\alpha }
=\braket{\alpha| \varPhi_{\rm I}}/\sqrt{\braket{\alpha}\braket{\varPhi_{\rm I}}}$ and matrix element $B_{\alpha\beta} = \bra{\beta} \hat\Psi^{}_\downarrow(0) \ket{\alpha}/\sqrt{\braket{\alpha} \braket{\beta}}$.
The sum rule of $A_\alpha$ and $B_{\alpha\beta}$ are
\begin{equation}
 \sum_{\alpha} |A_{\alpha}|^2=1, ~~~ L\sum_{\beta} |B_{\alpha\beta}|^2=1,
 \label{es:sum-rule}
\end{equation}
respectively, and $|A_\alpha|^2$ ($|B_{\alpha\beta}|^2$)  is  the weight of eigenstate $\ket{\alpha}$ in the overlap (density matrix element).

%
%
%
Using the eigenstates of Hamiltonian (\ref{es:H}), we can   calculate the eigenvalues of the Hamiltonian,  the overlap $A_\alpha$ and the matrix elements $B_{\alpha\beta}$ in terms of determinant representation of the norms and form factors.
Consequently, we may obtain the evolutions of the momentum and momentum distributions.
In particular,  guided by the sum rules, we can select the  microscopic states with
 large sum rule  weights that essentially comprise the oscillation features of the QF and revival  dynamics.
We give in details the calculations of the above mentioned quantities in next sections.

\section{S3.  Method for calculating  time evolution of impurity momentum}

\subsection{S3.1 Selection of the  eigenstates for quantum flutter}

In the BA equations Eqs. (2) in the main text, $\boldsymbol{I}$ and $\boldsymbol{J}$ denote  the quantum numbers of the charge and spin degrees of freedom, respectively, where $\boldsymbol{I}=\boldsymbol{I}_{N}=\{I_1,I_2,\cdots,I_N\}$, $\boldsymbol{J}=\boldsymbol{J}_M=\{J_1,J_2,\cdots,J_M\}$ and $M$ is the number of  spin-down particles.
For a given set of quantum numbers $\{\boldsymbol{I}_N, \boldsymbol{J}_M\}$, the  BA equations uniquely determine the wave numbers and spin rapidities  $\{k_1,k_2,\cdots,k_N; \lambda_1, \lambda_2, \cdots, \lambda_M\}$.
Consequently, one  BA solution/BA highest weight state
gives  $N-2M+1$ eigenstates, $\ket{\boldsymbol{I}_N, \boldsymbol{J}_M,\ell}=(\hat S^-)^\ell \ket{\boldsymbol{I}_N, \boldsymbol{J}_M,0}$, where $\ell=0,1,2,\cdots,N-2M$.
Here we denote $\ket{\boldsymbol{I}_N, \boldsymbol{J}_M,0}$ as the 
highest weight state, i.e., $\hat S^+\ket{\boldsymbol{I}_N, \boldsymbol{J}_M,0}=0$.
The states with non-zero values of the $\ell$ are non-highest weight states.
In the above, we defined the spin operators  $\hat S^-=\int {\rm d}x \hat \Psi^\dag_\downarrow(x) \hat \Psi_\uparrow(x)$ and $\hat S^+=\int {\rm d}x \hat \Psi^\dag_\uparrow(x) \hat \Psi_\downarrow(x)$.
The total spin $S$ and its projection  in  $z$-direction $S^{\rm z}$ are good quantum numbers of the state $\ket{\boldsymbol{I}_N, \boldsymbol{J}_M,\ell}$, namely,
\begin{equation}
 S=N/2-M,~~~S^{\rm z}=N/2-M-\ell.
\end{equation}


 There are three  sets of complete eigenstates   $\{\ket{\alpha}\}$, $\{\ket{\alpha'}\}$ and $\{\ket{\beta}\}$ which were inserted in the calculation of impurity momentum $K_\downarrow(t)$ Eq. (\ref{es:kd-1}).
 The eigenstates  include all of the highest and non-highest weight ones.
Guided by the sum rules, we need to select  enough states to calculate the dynamical evolutions of the momentum  $K_\downarrow(t)$ and momentum distributions.
Without losing accuracy, the following selection rules were  used to essentially simplify our numerical task:
\begin{itemize}
 \item[(i)]  The total  particle number is a good quantum number of the initial state $\ket{\varPhi_{\rm I}}$, such that
 $\{\ket{\alpha}\}$ and $\{\ket{\alpha'}\}$ consist of the states $\ket{\boldsymbol{I}_N,\boldsymbol{J}_M,\ell}$ with total particle number $N$.
 However,  the state  $\{\ket{\beta}\}$ in Eq. (\ref{es:kd-1}) must be the  state $\ket{\boldsymbol{I}_{N-1},\boldsymbol{J}_{M'},\ell}$ with the total particle number $N-1$, respectively.

 \item[(ii)]  The total spin is not a good quantum number of the initial state $\ket{\varPhi_{\rm I}}$, while $S^{\rm z}$ is a good quantum number, $\hat S^{\rm z}\ket{\varPhi_{\rm I}}=(N/2-1)\ket{\varPhi_{\rm I}}$.
 Together with the selection rule (i), the possible states of $\{\ket{\alpha}\}$ and $\{\ket{\alpha'}\}$ are $\ket{\boldsymbol{I}_N,J=\boldsymbol{J}_1,0}$ and $\ket{\boldsymbol{I}_N,\boldsymbol{J}_0,1}$.
 Whereas the  state $\{\ket{\beta}\}$  relates to the state  $\ket{\boldsymbol{I}_{N-1},\boldsymbol{J}_0,0}$,  where $\boldsymbol{J}_0$ is an empty set.

 \item[(iii)] When the impurity wave function $\phi_\downarrow(x)$ is a plane wave with a fixed momentum $Q$, the total momentum is also a good quantum number of $\ket{\varPhi_{\rm I}}$, $\hat K\ket{\varPhi_{\rm I}} = Q \ket{\varPhi_{\rm I}}$, so that $\brakets{\boldsymbol{I},\boldsymbol{J}_0,1}{\varPhi_{\rm I}} =\brakets{\boldsymbol{I},J,0}{\varPhi_{\rm I}}=0$ when the quantum numbers do not satisfy $K=Q$ according to Eq. 
     (3) in the main text.
 We only need to calculate the states with $K_\alpha=K_{\alpha'}=Q$ in our study.
\end{itemize}
Based on these selection rules, we need to obtain  the states $\ket{\boldsymbol{I}_N,J,0}$, $\ket{\boldsymbol{I}_N,\boldsymbol{J}_0,1}$ and $\ket{\boldsymbol{I}_{N-1},\boldsymbol{J}_0,0}$ ($\boldsymbol{J}_0$ is defined in selection rule (ii)).
We will give these states  in the following study.


%
%
For the states with all $N$ particles spin-up, we give a set of quantum numbers $\boldsymbol{I}_N$, get a set of wave numbers $\{k_j\}$ from the BA equations (2) and find the wave function of this eigenstate to be
\begin{eqnarray}
\ket{\boldsymbol{I},\boldsymbol{J}_0,0}
&&=
\int {\rm d}\boldsymbol{x} \varPhi_0(\boldsymbol{x})
\hat{\Psi}^{\dagger}_{\uparrow}(x_1) \dots
\hat{\Psi}^{\dagger}_{\uparrow}(x_N)
\vert 0 \rangle,
\nonumber\\
\varPhi_0(\boldsymbol{x})
&&=
\frac{1}{\sqrt{N!}}
\sum_{{\cal P}}^{}
(-1)^{{\cal P}}
{\rm e}^{{\rm i}\sum_{j}x_j k_{{\cal P}_j}}
\nonumber\\
&&\times
\prod_{i<j}
[k_{{\cal P}_i}-k_{{\cal P}_j}+{\rm i}c{\rm sign}(x_j-x_i)],
 \label{es:wavefunc0}
\end{eqnarray}
where ${\cal P}$ are the permutations of $\{1,2,\cdots,N\}$.
The total spin of this state is $S=S^{\rm z}=N/2$.
In fact, $\ket{\boldsymbol{I},\boldsymbol{J}_0,0}$ is the eigenstate of the Lieb-Liniger model.

There are two kinds of eigenstates with one spin-down particle, the highest weight states $\ket{\boldsymbol{I},J,0}$ and the non-highest weight states $\ket{\boldsymbol{I},\boldsymbol{J}_0,1}$.
For the highest weight state, a given set of quantum numbers $\{\boldsymbol{I}_N,J\}$ determines  a unique solution of the BA equations (2), namely, the wave numbers and spin rapidity $\{k_1,k_2,\cdots,k_N; \lambda \}$.
Then we can have explicit forms of  different  wave functions. The highest weight state is given by
\begin{eqnarray}
\ket{\boldsymbol{I},J,0}
&&=
\int {\rm d}\boldsymbol{x} \varPhi_1(\boldsymbol{x})
\hat{\Psi}^{\dagger}_{\downarrow}(x_1) \dots
\hat{\Psi}^{\dagger}_{\uparrow}(x_N)
\vert 0 \rangle,
\nonumber
\\
\varPhi_1(\boldsymbol{x})
&&=
\sum_{l=1}^{N}
\frac{1}{\sqrt{N!}}
\bigg[
\sum_{{\cal P}}^{}
(-1)^{{\cal P}}
{\rm e}^{{\rm i}\sum_{j}x_jk_{{\cal P}_j}}
\nonumber
\\
&& \times
\prod_{i<j}^{N}
[k_{{\cal P}_i}-k_{{\cal P}_j}+{\rm i}c~{\rm sign}(x_j-x_i)]
\nonumber
\\
&& \times
\prod_{j \neq l}^{}
\Big[\lambda-k_{{\cal P}_j}+{\rm i}\frac{c}{2}{\rm sign}(x_l-x_j)\Big]
\bigg].
 \label{es:wavefunc1}
\end{eqnarray}
The total spin of this state is $S=S^{\rm z}=N/2-1$.
Using the relation Eq. (\ref{es:wavefunc0}), $\ket{\boldsymbol{I},\boldsymbol{J}_0,1}=\hat S^-\ket{\boldsymbol{I},\boldsymbol{J}_0,0}$, the non-highest weight states  is given by
\begin{eqnarray}
\ket{\boldsymbol{I},\boldsymbol{J}_0,1}
 &&
 =
 \sum_{l=1}^N
\int {\rm d}\boldsymbol{x} \varPhi_0(\boldsymbol{x})
\nonumber\\
&& \times
\hat{\Psi}^{\dagger}_{\uparrow}(x_1)
\dots
\hat{\Psi}^{\dagger}_{\downarrow}(x_l) \dots
\hat{\Psi}^{\dagger}_{\uparrow}(x_N)
\vert 0 \rangle.
\label{es:wavefunc2}
\end{eqnarray}
The total spin of this state $S = N/2$ and $S^{\rm z} = N/2-1$.

\subsection{S3.2 Matrix element}

Based on the discussions above,  we need to calculate  $A_\alpha$ and $B_{\alpha,\beta}$  for the time evolution of impurity momentum $K_\downarrow(t)$.
Using  the specific forms of the wave functions of the relevant  states Eqs. (\ref{es:wavefunc0}-\ref{es:wavefunc2}), and following   the method \cite{cs:BPozsgay2012JPA,cs:BvandenBerg2016PRL2}.
we  can directly calculate $A_\alpha$ and $B_{\alpha,\beta}$.
Explicitly, we have
\begin{eqnarray}
  A_{\alpha }
  && = \frac{\braket{\alpha|\varPhi_{\rm I}}}{\sqrt{\braket{\alpha}\braket{\varPhi_{\rm I}}}}
   =\int {\rm d}x  \frac{\bra{\alpha}\phi_\downarrow(x) \Psi^\dag_\downarrow(x) \ket{\varOmega}}{\sqrt{\braket{\alpha}\braket{\varPhi_{\rm I}}}}
  \nonumber\\
  && =\int {\rm e}^{-{\rm i} K_\alpha x} {\rm d}x  \phi_\downarrow(x) \frac{\bra{\alpha} \Psi^\dag_\downarrow(0)  \ket{\varOmega}}{\sqrt{\braket{\alpha}\braket{\varPhi_{\rm I}}}},
  \\
  \braket{\varPhi_{\rm I}}
  &&=
\int {\rm d} y  \phi_{\downarrow}^{*}(y)
\bra{\varOmega} \hat{\Psi}_{\downarrow}^{}(y)
\int {\rm d} x  \phi_{\downarrow}(x) \hat{\Psi}_{\downarrow}^{\dagger}(x)
\ket{\varOmega}
\nonumber \\
&&=
\int {\rm d} x  |\phi_{\downarrow}(x)|^2
\braket{\varOmega}.
\end{eqnarray}
We further  calculate norms and overlaps, where $\ket{\varOmega} = \ket{\boldsymbol{I}_{N-1},\boldsymbol{J}_0,0}$, $\ket{\alpha} = \ket{\boldsymbol{I}_{N},J,0}$ or $\ket{\alpha} = \hat S^-\ket{\boldsymbol{I}_{N},\boldsymbol{J}_0,0}=\ket{\boldsymbol{I}_{N},\boldsymbol{J}_0,1}$.

The norm of the state $\ket{\boldsymbol{I},\boldsymbol{J}_0,0}$ is given by
\begin{eqnarray}
&&\braket{\boldsymbol{I}_N,\boldsymbol{J}_0,0}
=
\prod_{i<j} [(k_i-k_j)^2+c^2] {\rm det}(\mathcal{G}),
\label{es:norm-1-N-up}
\\
&&\mathcal{G}_{ij}
=
\delta_{i,j}\Big[L+\sum_{l=1}^{N}
\phi_{1}(k_i-k_l)\Big]-\phi_{1}(k_i-k_j),
\nonumber\\
&&\phi_{n}(u)
= \frac{2cn}{n^2u^2+c^2},
\nonumber
\end{eqnarray}
where $\{k_1,k_2,\cdots,k_N\}$ are the solution of BA equation (2) with the quantum numbers $\boldsymbol{I}_N$.
The norm of non-highest weight state $\ket{\boldsymbol{I},\boldsymbol{J}_0,1}$ can also be calculated by using Eq. (\ref{es:norm-1-N-up}), namely,
\begin{equation}
\braket{\boldsymbol{I}_N,\boldsymbol{J}_0,1}
=\bra{\boldsymbol{I}_N,\boldsymbol{J}_0,0}\hat{S}^{+}\hat{S}^{-}\ket{\boldsymbol{I}_N,\boldsymbol{J}_0,0}
=N\braket{\boldsymbol{I}_N,\boldsymbol{J}_0,0}.
\nonumber
\end{equation}
Then the norm of the state $\ket{\boldsymbol{I},J,0}$ is given by the following equation
\begin{eqnarray}
\braket{\boldsymbol{I},J,0}
&&=
\bigg\vert
\frac{1}{-{\rm i}c}
\prod_{j=1}^{N}
[\lambda-k_j-{\rm i}c']
\prod_{i<j}^{}
[k_i-k_j+{\rm i}c]
\bigg\vert^2
\nonumber\\
&&\times
c\ {\rm det}\mathcal{J},
\label{es:norm-2-one-down}
\end{eqnarray}
where ${\cal J}$ is a $N+1$-dimensional matrix, explicitly,
\begin{eqnarray}
&& \mathcal{J} =
\begin{pmatrix}
J_{kk}        & J_{k\lambda} \\
J_{\lambda k} & J_{\lambda \lambda} \\
\end{pmatrix}_{N+1},
\nonumber \\ &&
(J_{kk})_{ij}
=
\delta_{ij}\Big[L+\sum_{m=1}^{N}
\phi_{1}(k_i-k_m)-\phi_{2}(k_i-\lambda)\Big]
\nonumber\\  && \hspace{20pt}
-\phi_{1}(k_i-k_j),
\nonumber \\ &&
(J_{k\lambda})_{i,N+1}
=
\phi_{2}(k_i-\lambda),~~
(J_{\lambda k})_{N+1,j}
=-\phi_{2}(k_j-\lambda),
\nonumber \\ &&
(J_{\lambda \lambda})_{N+1,N+1}
=
\sum_{m=1}^{N}
\phi_{2}(k_m-\lambda), \nonumber
\end{eqnarray}
and $\{k_1,k_2,\cdots,k_N; \lambda\}$ are the solution of BA equation (2) with the quantum numbers $\boldsymbol{I}_N$ and $J$.

To calculate $\bra{\alpha} \Psi^\dag_\downarrow(0)  \ket{\varOmega}$ we need the matrix elements  $\bra{\boldsymbol{I}'_{N-1},\boldsymbol{J}_0,0}$$\hat \Psi_\downarrow(0)$$\ket{\boldsymbol{I}_{N},J,0}$ and $\bra{\boldsymbol{I}'_{N-1},\boldsymbol{J}_0,0} \hat \Psi_\downarrow(0)$ $\ket{\boldsymbol{I}_{N},\boldsymbol{J}_0,1}$ for the highest and non-highest weight $\ket{\alpha}$, respectively.
The matrix element $\bra{\boldsymbol{I}'_{N-1},\boldsymbol{J}_0,0}$ $\hat \Psi_\downarrow(0)\ket{\boldsymbol{I}_{N},J,0}$ is give by
\begin{eqnarray}
&&
\bra{\boldsymbol{I}'_{N-1},\boldsymbol{J}_0,0} \hat \Psi_\downarrow(0)\ket{\boldsymbol{I}_{N},J,0}
={\sqrt{N}(N-1)!} {\rm det} \mathcal{M}
\nonumber\\
&& \hspace{20pt} \times
\frac
{\prod_{i>j}(k_i-k_j+{\rm i}c)}
{\prod_{l>m}(q_l-q_m+{\rm i}c)}
\frac
{-{\rm i}c}
{\prod_{j}(\lambda-k_j-{\rm i}c')}.
\label{es:overlap-1}
\end{eqnarray}
Here the $(N-1)\times(N-1)$ matrix $\mathcal{M}$ has elements $\mathcal{M}_{jk}=M_{jk}-M_{N,k}$,
\begin{eqnarray}
&& M_{jk} =
t(q_k-k_j) h_2(\lambda-k_j)
\frac
{\prod_{m=1}^{N-1}h_1(q_m-k_j)}
{\prod_{m=1}^{N}h_1(k_m-k_j)}
\nonumber\\
&& \hspace{20pt}+
t(k_j-q_k) h_2(k_j-\lambda)
\frac
{\prod_{m=1}^{N-1}h_1(k_j-q_m)}
{\prod_{m=1}^{N}h_1(k_j-k_m)}, \nonumber \\
&&
h_{n}(u) =u+{\rm i}\frac{c}{n}, ~~~~
t(u)=\frac{-c}{u(u+{\rm i}c)}, \nonumber
\end{eqnarray}
where $\{q_1,q_2,\cdots,q_{N-1} \}$ are the solution of the BA equations (2) with the quantum numbers $\boldsymbol{I}_{N-1}$.

For the matrix element of the non-highest weight state $\ket{\alpha}$, we have
\begin{eqnarray}
 &\bra{\boldsymbol{I}'_{N-1},\boldsymbol{J}_0,0} \hat \Psi_\downarrow(0)\ket{\boldsymbol{I}_{N},\boldsymbol{J}_0,1}
 \nonumber\\
 &=\bra{\boldsymbol{I}'_{N-1},\boldsymbol{J}_0,0} \hat \Psi_\downarrow(0) \hat S^-\ket{\boldsymbol{I}_{N},\boldsymbol{J}_0,0}
 \nonumber\\
 &=N\bra{\boldsymbol{I}'_{N-1},\boldsymbol{J}_0,0} \hat{\Psi}_{\uparrow}(0) \ket{\boldsymbol{I}_{N},\boldsymbol{J}_0,0},
 \nonumber
\end{eqnarray}
where $\bra{\boldsymbol{I}'_{N-1},\boldsymbol{J}_0,0} \hat{\Psi}_{\uparrow}(0) \ket{\boldsymbol{I}_{N},\boldsymbol{J}_0,0}$ is the  matrix element of the Lieb-Liniger model
\begin{eqnarray}
&&
\bra{\boldsymbol{I}'_{N-1},\boldsymbol{J}_0,0} \hat{\Psi}_{\uparrow}(0) \ket{\boldsymbol{I}_{N},\boldsymbol{J}_0,0}
\nonumber \\
&&=
(N-1)!\sqrt{N}
\frac
{\prod_{i>j}(k_i-k_j+{\rm i}c)}
{\prod_{l>m}(q_l-q_m+{\rm i}c)}
{\rm det} \mathcal{S},
\label{es:overlap-2}
\end{eqnarray}
where $\mathcal{S}_{i,j} = S_{i,j}-S_{N,j}$,
\begin{eqnarray}
&&S_{ij}
=
t(q_j-k_i)
\frac
{\prod_{m=1}^{N-1}h_1(q_m-k_i)}
{\prod_{m=1}^{N}h_1(k_m-k_i)}
\nonumber \\
&&\hspace{20pt}
-t(k_i-q_j)
\frac
{\prod_{m=1}^{N-1}h_1(k_i-q_m)}
{\prod_{m=1}^{N}h_1(k_i-k_m)},
\nonumber\\
&&h_{n}(u)
=u+{\rm i}\frac{c}{n},~~
t(u)=\frac{-c}{u(u+{\rm i}c)}.
\nonumber
\end{eqnarray}
The above determinant forms are convenient for us to perform numerical calculations.



 %
 %
 %
 %

In order to calculate $B_{\alpha,\beta}$
\begin{equation}
 B_{\alpha\beta}=\frac{\bra{\beta} \hat{\Psi}_{\downarrow}^{}(0) \ket{\alpha} }{ \sqrt{\braket{\alpha} \braket{\beta}}},
\end{equation}
we need to calculate norms $\braket{\alpha}$, $\braket{\beta}$ and the overlap $\bra{\beta} \hat{\Psi}_{\downarrow}^{}(0) \ket{\alpha}$.
Where $\ket{\beta} = \ket{\boldsymbol{I}'_{N-1},\boldsymbol{J}_0,0}$ and is the Bethe state of $N-1$ particles with all spin up.
Similar to the calculation of $A_\alpha$, here $\ket{\alpha}$ also involves the  highest or non-highest weight states, namely, $\ket{\alpha} = \ket{\boldsymbol{I}_{N},J,0}$ and $\ket{\alpha}=\ket{\boldsymbol{I}_{N},\boldsymbol{J}_0,1}$, respectively.
The norms  can be calculated by Eqs. (\ref{es:norm-1-N-up}, \ref{es:norm-2-one-down}).
Similarly,  for the highest weight state $\ket{\alpha}$ we can calculate the matrix element by using Eq. (\ref{es:overlap-1})
\begin{eqnarray}
&&
\bra{\beta}\hat{\Psi}_{\downarrow}(0)\ket{\alpha}
=\bra{\boldsymbol{I}'_{N-1},\boldsymbol{J}_0,0} \hat{\Psi}_{\downarrow}(0) \ket{\boldsymbol{I}_{N},J,0}. \nonumber
\end{eqnarray}
When $\ket{\alpha}$ is the non-highest weight state
\begin{equation}
\begin{split}
\bra{\beta}\hat{\Psi}_{\downarrow}(0)\ket{\alpha}
&
=
\bra{\boldsymbol{I}'_{N-1},\boldsymbol{J}_0,0} \hat{\Psi}_{\downarrow}(0) \hat{S}^{-} \ket{\boldsymbol{I}_{N},\boldsymbol{J}_0,0}\\
&
=
N\bra{\boldsymbol{I}'_{N-1},\boldsymbol{J}_0,0} \hat{\Psi}_{\uparrow}(0) \ket{\boldsymbol{I}_{N},\boldsymbol{J}_0,0}
,\nonumber
\end{split}
\end{equation}
which  is given by Eq. (\ref{es:overlap-2}).

\begin{figure*}[t]
	\renewcommand{\thefigure}{s1}
	\centering
	\begin{center}
		\includegraphics[width=1\linewidth]{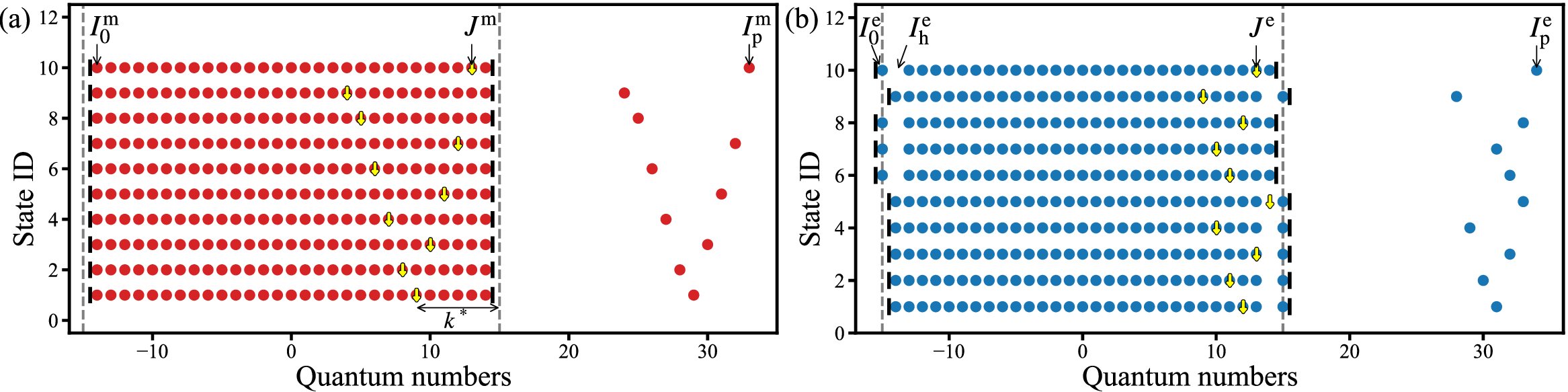}
	\end{center}
	\caption{
		Quantum numbers of magnon-like states (a) and exciton-like states (b).
		The dots and yellow arrows are quantum numbers $\boldsymbol{I}$ and $J$, respectively.
		The short thick straight lines show the clear ``Fermi surfaces'' structure of $\boldsymbol{I}$.
		The thin black dashed straight lines are the Fermi points of the system where $I=\pm N/2$.
		We select the top 10 states with largest $|A|^2$ of magnon- and particle-hole excitation
 states and plot them in (a) and (b), respectively.
When the hole in (b) locates  in deep ``Fermi sea'', we name the state as exciton-like in the paper.
		Here $\gamma=10$, $Q=1.33 k_{\rm F} $ and $N=30$.
	}
	\label{fs:magnon-exciton}
\end{figure*}

\subsection{S3.3 Magnon- and exciton-like states}

	Quasiparticles are used to describe individual or collective excitations.
	Taking the magnon as an example, its appearance is due to the transmission of spin wave in a ferromagnetic state.
	The essence of excitation is a change in quantum numbers in Bethe ansatz.
	It describes quasi-particles in this way. Using the ground state as a reference, $-N/2<\boldsymbol{I}<N/2$.
	One type of excited states which  is formed after impurity enters the system is worth noting.
	As shown in (a) of FIG. 2 [as mentioned in the main text], the red states have a complete Fermi Sea (without hole inside), plus an external particle $I_{\rm p}$.
	The quantum number $J$ indicates  that the system undergoes a spin flip and propagates at a certain speed.
	The excitation energy  is $E \approx \frac{2\pi^2}{3\gamma m}q^2$ for  a strong interaction, showing a magnon-like state.
	Here $q$ is the total momentum of the system.

	In contrast, for the exciton states (with one hole in the ``Fermi sea''),  their  $J$s of  the blue states  in Fig. 2 in the main text  is close to $-N/2$, resulting in a spin excitation energy is about $0$.
	We   will further  demonstrate  in the following Section 4,  a particle at the left Fermi point coupled to a deep hole in Fermi sea form the exciton state.
	This situation is much  like an exciton, where an electron in the conduction band is bound to a hole in the valence band.
	Therefore  we call the blue states in  Fig. 2 in the main text  as exciton-like states.
	After quenching, the system undergoes dynamic evolution through energy transfer between different  states.

	For an impurity with variation of the mass and the initial momentum, the quenching dynamics results in the excitation of particle-hole pairs, where the single particle-hole pairs are the primary contributors. The closer the hole is to the Fermi surface, the higher the weight of the state is \cite{Burovski:2014,Gamayun:2018,Gamayun:2020,Caux:2020,Gamayun:2023}, where the impurity initial velocity is not larger than the sound velocity of the medium. 
In the papers \cite{Burovski:2014,Gamayun:2018,Gamayun:2020,Caux:2020},  authors gave insightful understanding of the long time evolution from the particle-hole excitations. In the long time limit, diagonal ensemble plays an important role so that the overlaps between the initial state and the excited states are the key ingredient for their study of the dynamics of the impurity of the impurity model. 
 However, for a supersonic  impurity, this case is not the full features of the impurity dynamics. 
	
	Here we use quantum numbers $\boldsymbol{I}$ to classify the highest weight states.
	In principle, all the possible choices of quantum numbers satisfy the restrictions of Eq. (2) [as mentioned in the main text] and the selection rules above should be taken account.
	Here we only classify the states where $\boldsymbol{I}$ have a clear structure of ``Fermi sea'',
	there is no hole or only one hole $I_{\rm h}$ inside the ``Fermi see'' and only one particle $I_{\rm p}$ outside the ``Fermi sea''.

The magnon- and exciton-like states essentially comprise the feature of   QF and quantum revival phenomena, see FIG.2 in the main text.
The dots and  arrows denote  quantum numbers $\boldsymbol{I}$ and $J$, respectively.
We regard  the state without hole inside the ``Fermi sea'' as magnon-like state, FIG. \ref{fs:magnon-exciton} (a), while  the state with only one hole in the deep Fermi sea  as exciton-like state, FIG. \ref{fs:magnon-exciton} (b).
The $\boldsymbol{I}$ of magnon- and exciton-like states have the following form
\begin{eqnarray}
 \begin{split}
  &\boldsymbol{I}^{\rm m}=\{I^{\rm m}_0,I^{\rm m}_1,\cdots,I^{\rm m}_{N-2},I^{\rm m}_{\rm p}\},
 \\
  &\boldsymbol{I}^{\rm e}=\{I^{\rm e}_0,I^{\rm e}_1,\cdots,I^{\rm e}_{h-1},I^{\rm e}_{h+1},\cdots,I^{\rm e}_{N-1},I^{\rm e}_{\rm p}\},
 \end{split}
\end{eqnarray}
respectively.
We denote the quantum number of spin-down particle as $J^{\rm m}$ and $J^{\rm e}$ for magnon- and exciton-like states, respectively.
Here, $I_\ell$ are the quantum numbers in the ``Fermi sea'', $I_\ell = I_0+\ell$
, $I_0$ is the starting quantum number of the ``Fermi sea", $\ell=1,2,\dots,N-1$
, $I_{\rm p}$ is the quantum number of particle excitation and $I^{\rm e}_h$ is location of hole in the exciton-like states.
Note that the quantum number $J$ is fixed for a given $\boldsymbol{I}$ of the emitted particle  when the impurity is injected with a large moment $Q$ into the medium of the Lieb-Liniger Bose gas.
This is mainly because of the conservation of momentum
\footnote{In the TABLE \ref{ts:weights}, the sum rule of magnons (excitons) also involves the magnons with other $I^{\rm m}_0$ ($I^{\rm e}_0$).}.
We denote the most important quantum numbers  in the study of  the QF and quantum revival phenomena, namely,
\begin{equation}
 \begin{array}{ll}
  \mbox{magnon-like states: } & \{I^{\rm m}_{\rm p},J^{\rm m}\},\\
  \mbox{exciton-like states: } & \{I^{\rm e}_{\rm p},I^{\rm e}_{\rm h},J^{\rm e}\}.
 \end{array}
  \label{es:mag-exc-quan-num}
\end{equation}

\begin{table}[t]
\caption{
Values of sum rule $\sum |A_\alpha|^2$ and state numbers in our numerical calculations.
The values outside of brackets are $\sum |A_\alpha|^2$ and the values inside of brackets are numbers of states taken account in our calculations. Here, $N=30$ and $\gamma=10$.}
\begin{ruledtabular}
\begin{tabular}{c|c|ccccc}
  $Q$         & total     & magnon &exciton&other HS$^{*1}$ & NHS$^{*2}$\\
\hline
$1.33k_{\rm F}$ &0.97(86795) &0.711 &0.040 &0.192 &0.025   \\
$1.07k_{\rm F}$ &0.96(89928) &0.696 &0.041 &0.193 &0.030   \\
$1.00k_{\rm F}$ &0.95(90682) &0.691 &0.041 &0.187 &0.031   \\
$0.80k_{\rm F}$ &0.95(92817) &0.673 &0.043 &0.201 &0.033   \\
$0.53k_{\rm F}$ &0.95(95190) &0.641 &0.047 &0.223 &0.039   \\
$0.13k_{\rm F}$ &0.95(97221) &0.539 &0.074 &0.263 &0.074   \\
 \end{tabular}
\begin{flushleft}
(*1), other HS: other highest weight states;\\
(*2), NHS: non-highest weight states.
\end{flushleft}
\end{ruledtabular}
\label{ts:weights}
\end{table}

Now we have all ingredients to calculate precisely  $K_\downarrow(t)$  to capture the essence of the dynamics of the  QF and quantum revival.
Without losing generality, we in this paper take the impurity wave packet as a plane wave.
Precisely speaking, we also treat the  sum rule $\sum |A_\alpha|^2$  larger than $95\%$ in our actual calculations,  see TABLE. \ref{ts:weights}.
As shown in TABLE. \ref{ts:weights}, the magnon-like states have the largest $\sum |A_\alpha|^2$ and they are the most important states in the supersonic impurity phenomenon.
The sum rule of exciton-like states are relatively small.
We will show the importance of these states in the QF phenomenon in Sec. S4 in this supplemental material.
The states other than the mentioned  highest weight states and the  non-highest weight states
are all  necessary  in the calculation of $K_\downarrow(t)$.
Although
they make very 
small  contributions to  the dynamics of the QF and quantum revival, 
our calculations show that if the contribution of these states is ignored, the results will have obvious difference (about 5 percent), indicating that these states have non-trivial contribution.

The sum rule of the matrix element $B_{\alpha\beta}$ depends on $\alpha$, see Eqs.(\ref{es:sum-rule}).
Based on  the sum rule of weights, we observe that  the magnon-like states are of the most importance in the matrix element $B_{\alpha\beta}$.
We denote $N_{\rm AMS}$ as  the number of accounted magnon-like states  (AMS) in our numerical calculations.
In this paper, we request the numerical sum rule $L\sum_{\alpha \in{\rm AMS}}\sum_{\beta}|B_{\alpha\beta}|^2>0.97 N_{\rm AMS}$.

\subsection{S3.4 Exciton energy in  the thermodynamic limit}

In this section, we discuss the excitation energies of magnon- and exciton-like states in  the thermodynamic limit.
Building on the BA solution in  the thermodynamic limit, i.e., $N\to\infty$, $L\to\infty$ and $\gamma=cL/N$ is finite.
The energy of excited states is calculated by using the
thermodynamic Bethe ansatz (TBA) equations \cite{cs:SJGu2002IJMPB, cs:Li-YQ:2003, cs:Guan-Batchelor-Takahashi}.
The medium $\ket{\varOmega}$ is the ground state of the Lieb-Liniger gas and the TBA equations of this model  is given in \cite{cs:Lieb-Liniger, cs:CNYang1969JMP, cs:Guan2015CPB}, namely,
\begin{equation}
 \label{es:TBAE-c}
 \begin{split}
 &
 \rho_{\rm c}(k) + \rho_{\rm c}^{\rm h}(k) = \frac{1}{2\pi}+
 \int_{-k_0}^{k_0} a_2(k-k') \rho^c(k') {\rm d} k',
 \\
 &\varepsilon_{\rm c}(k) = k^2-\mu+ \int_{-k_0}^{k_0} a_2(k-k') \varepsilon_{\rm c}(k'){\rm d}k',
 \end{split}
\end{equation}
where $\rho_{\rm c}(k)$ is the linear density,
\begin{equation}
 \rho_{\rm c}(k) = \left\{
 \begin{array}{ll}
  \rho_{\rm c}(k), & |k|<k_0,\\
  0, & |k|>0,
 \end{array}
 \right.
 \rho_{\rm c}^{\rm h}(k) = \left\{
 \begin{array}{ll}
  0, & |k|<k_0,\\
  \rho_{\rm c}^{\rm h}(k), & |k|>0,
 \end{array}
 \right.
 \nonumber
\end{equation}
$\varepsilon_{\rm c}(k)$ is the dressed energy of the charge sector and the integral kernal $a_n(x)=nc/[2\pi (x^2+n^2c^2)]$.
Here, $k_0$ is the Fermi point (cut-off) of the wave numbers $k$ and it is determined by $\int_{-k_0}^{k_0} \rho(k) {\rm d}k=N/L$.
$\mu$ is the chemical potential and it is determined by the condition $\varepsilon_{\rm c}(k_0)=0$.
%
%
The TBA equations  for the  density and dressed energy of the spin degree of freedom are given by
\begin{equation}
 \label{es:TBAE-s}
 \begin{split}
 & \rho_{\rm s}(\lambda) + \rho_{\rm s}^{\rm h}(\lambda)=  \int_{-k_0}^{k_0} a_1(k-\lambda)\rho_{\rm c}(k){\rm d}k,\\
 & \varepsilon_{\rm s}(\lambda)= - \int_{-k_0}^{k_0} a_1(k-\lambda)\varepsilon_{\rm c}(k){\rm d}k,
 \end{split}
\end{equation}
respectively.
%

\begin{figure}[t]
\begin{center}
 \includegraphics[width=0.7\linewidth]{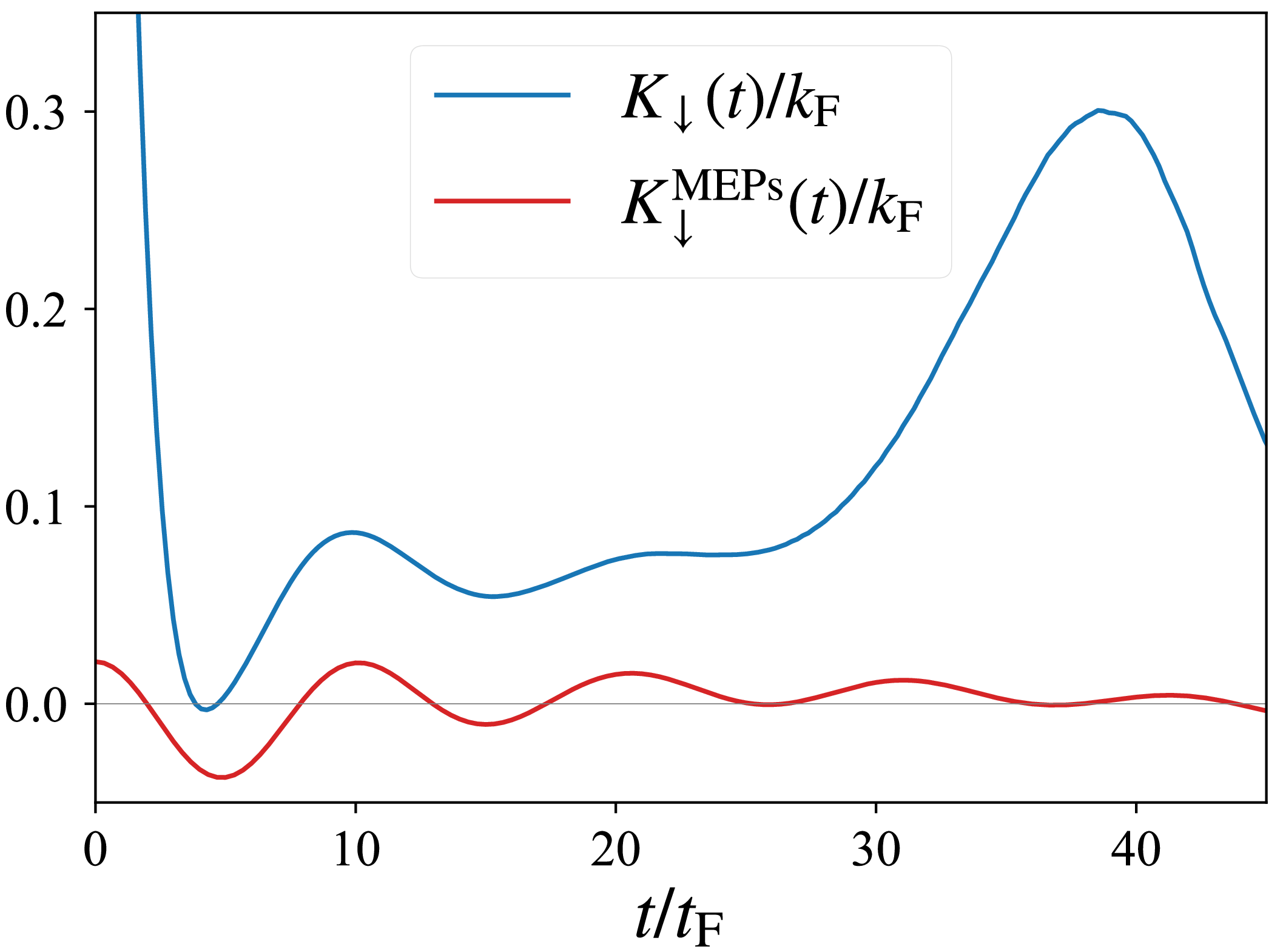}
\end{center}
 \caption{
 The time evolution of impurity momentum $K_\downarrow(t)$ when $\gamma=10$, $Q=1.33k_{\rm F}$ and $N=30$.
 The blue line is calculated with all of the states in TABLE. \ref{ts:weights} and
 the red line is calculated by the special MEPs in Eq. (\ref{es:MEP-quan-num}).
}
\label{fs:K-t}
\end{figure}

The starting quantum number $I_0$ of the magnon- and exciton-like states in Eqs. (\ref{es:mag-exc-quan-num}) are near by the left Fermi point, $\lim_{L\to \infty} 2\pi I_0/L = -k_{\rm F}$, where
the Fermi momentum $k_{\rm F}=\pi N/L$.
The excitation energies carried by the quantum numbers near by the Fermi surface are zero.
The excitation energies of magnon-like and exciton-like states  associated with the quantum numbers  Eqs. (\ref{es:mag-exc-quan-num}) can be expressed as \cite{cs:Li-YQ:2003, cs:Guan2015CPB}
\begin{equation}
 \label{es:excitation-energy}
 \begin{split}
  \Delta E_{\rm m}
 & =\mu+ \varepsilon_{\rm c}(k^{\rm m}_{\rm p})+\varepsilon_{\rm s}(\lambda^{\rm m}),
 \\
 \Delta E_{\rm e}
 & =\mu+ \varepsilon_{\rm c}(k^{\rm e}_{\rm p})-\varepsilon_{\rm c}(k^{\rm e}_{\rm h})
 +\varepsilon_{\rm s}(\lambda^{\rm e}),
 \end{split}
\end{equation}
respectively.
Here, $k^{\rm m,e}_{\rm p}$, $k^{\rm e}_{\rm h}$ and $\lambda^{\rm m,e}$ are the rapidities of the corresponding quantum numbers $I^{\rm m,e}_{\rm p}$, $I^{\rm e}_{\rm h}$ and $J^{\rm m,e}$, respectively.
They  can be determined by the following equations
\begin{equation}
 \begin{split}
 \frac{I}{L}
 &= \int_0^{k} [\rho^{}_{\rm c}(k')+\rho^{\rm h}_{\rm c}(k')] {\rm d} k',
 \\
 \frac{J}{L}
 &= \int_0^{\lambda} [\rho^{}_{\rm s}(\lambda')+\rho^{\rm h}_{\rm s}(\lambda')] {\rm d} \lambda'.
 \end{split}
\end{equation}

\section{S4. Quantum flutter}

We presented  the expression of the impurity momentum  $K_\downarrow(t)$  in Sec. S2.
Using the determinant formula of the norms, overlap and matrix element obtained in the Sec. S3, we presented  the impurity momentum  $K_\downarrow(t)$ and momentum distributions in the main text.
%
In order to conceive  the microscopic origin  of QF, we first study  the frequency (energy) spectrum of $K_\downarrow(t)$
\begin{equation}
 \tilde K_{\downarrow}(E)
 = \frac{1}{2\pi} \int {\rm d} t  {\rm e}^{-{\rm i}E t} K_\downarrow(t).
 \label{es:K-E-0}
\end{equation}
In Eqs. (\ref{es:kd-1}), $K_{\alpha\alpha'}$ is the momentum matrix element of the state pair $\{\ket{\alpha},\ket{\alpha'}\}$ and it has close relation with frequency spectrum $\tilde K_\downarrow(E)$ Eq. (\ref{es:K-E-0}).
We can calculate $\tilde K_\downarrow(E)$ by taking the average value of $K_{\alpha\alpha'}$ in a small energy interval
\begin{equation}
 \tilde K_\downarrow(E) = \frac{L}{2\Delta E} \sum_{\alpha\alpha'}{}^{\Delta E} K_{\alpha\alpha'},
 \label{es:K-E}
\end{equation}
where $K_{\alpha\alpha'}$ is given by Eq. (\ref{es:kd-1}).
In the above equation, $\Delta E\ll E_{\rm F}$ and $E-\Delta E <E_\alpha-E_{\alpha'}<E+\Delta E$,  and the summation $\Sigma^{\Delta E}$ is taken  over all of the state pairs.
\begin{figure}[t]
\begin{center}
 \includegraphics[width=1\linewidth]{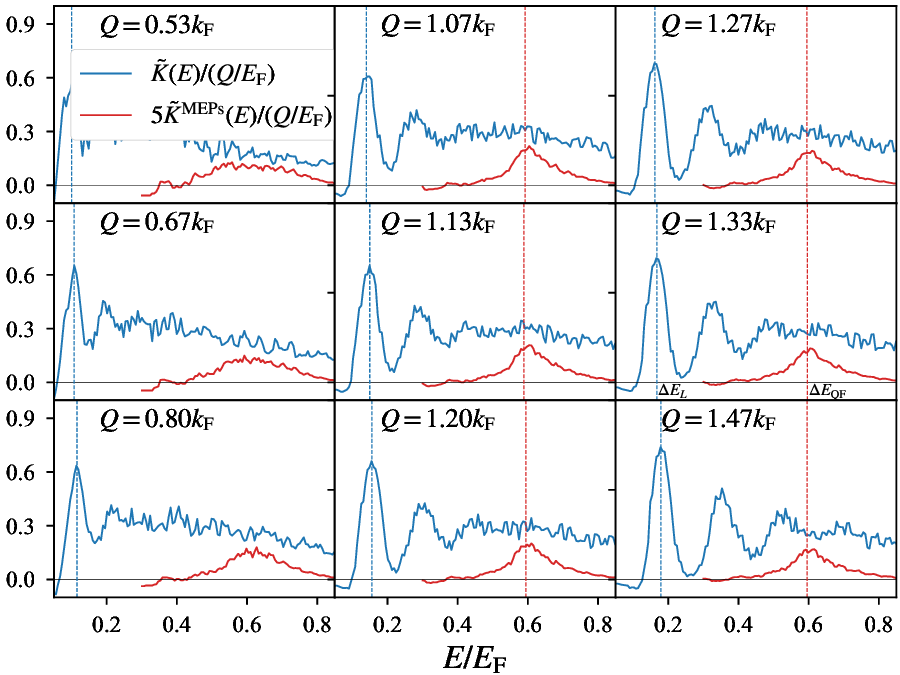}
\end{center}
\caption{ The impurity momentum $\tilde K_\downarrow(E)$ in frequency space  for different initial impurity momentum  $Q$ and a fixed interaction strength  $\gamma=10$.
The blue lines show  $\tilde K_\downarrow(E)$ in frequency space by  means of the Eq. (\ref{es:K-E}).
The vertical blue  lines indicates the revival frequency determined by the
magnon pairs
 with different values of the $Q$, see Sec. S5.
It is near the frequency  $0.15 E_{\rm F}$,  the typical energy  of quantum revivals, $\Delta E_L$.
The vertical red dashed lines nearby $0.6E_{\rm F}$ gives  the typical frequency  of the  QF, i.e.  $\Delta E_{\rm QF}$.
In the calculation of   $\tilde K_\downarrow(E)$ via Eq. (\ref{es:K-E}), we set $N=30$ and $\Delta E=0.03E_{\rm F}$.
 While the red lines are 5 times of the actual  value.
}
\label{fs:K-E}
\end{figure}

As being given in Eq. (\ref{es:K-E-0}), $\tilde K_\downarrow(E)$ is the oscillation amplitude of $K_\downarrow (t)$ at the frequency (energy) $E$.
We observe that the state pairs $\{\ket{\alpha},\ket{\alpha'}\}$ with an energy difference $E\sim E_\alpha- E_{\alpha'}$ essentially attribute to the oscillation nature of the  impurity momentum $\tilde K_\downarrow(E)$,  see Eq. (\ref{es:K-E}).
$\tilde K_\downarrow(E)$ is plotted  in FIG. \ref{fs:K-E} in which  several peaks were observed.
The first peak of $\tilde K_\downarrow(E)$ reveals the typical energy of quantum revival, which is governed  by the magnon pairs with nearest neighbour quantum numbers $I^{\rm m}_{\rm p}$.
 The second peak is from the magnon pairs with next nearest neighbour $I^{\rm m}_{\rm p}$, and so on.
We will discuss about the revival dynamics later.
The numerical result of $\tilde K_\downarrow(E)$ shows that the typical energy of QF $\Delta E_{\rm QF}$, is nearby $0.6E_{\rm F}$ for the system with  $\gamma=10$.
This strikingly indicates that  the frequency of the QF does not dependent on the initial momentum of the impurity once it is over the intrinsic sound velocity of the medium.
We also observed from  $\tilde K_\downarrow(E)$ that   the QF information of $\tilde K_\downarrow(E)$ is concealed by the peaks of magnon pairs, see  FIG. \ref{fs:K-E}.

So far, we realize that magnon pairs do not really contribute the frequency of  the  QF.
Such an oscillation feature of QF  is essentially resulted in from the magnon-exciton pairs (MEPs) described by the quantum numbers Eqs. (\ref{es:mag-exc-quan-num}).
We observe that  the quantum number  $I_{\rm p}$ in the two states of  one MEP are the same, presenting an  emitted particle.
Based on the conservation of the momentum, we only need to consider  the quantum numbers of hole in the exciton and spin-down  quantum number in magnon state
\begin{equation}
 \{I^{\rm e}_{\rm h},J^{\rm m}\}.
 \label{es:MEP-quan-num}
\end{equation}
The other quantum numbers can be given by  $I^{\rm e}_{\rm h},J^{\rm m}$, namely,
 $I^{\rm m}_{\rm p}=QL/2\pi+J^{\rm m}$ and $J^{\rm e}  = I^{\rm e}_{\rm p} - QL/2\pi-N/2- I^{\rm e}_{\rm h} = J^{\rm m} -N/2 -I^{\rm e}_{\rm h}$.
We take the summation in Eq. (\ref{es:K-E})   over the selected MEPs in Eq. (\ref{es:MEP-quan-num}) and denote it as
$\tilde K^{\rm MEPs}_\downarrow(E)$.
We plot  $\tilde K^{\rm MEPs}_\downarrow(E)$  (red lines)  in FIG. \ref{fs:K-E}.
It is clear seen that the QF oscillations of $K_\downarrow(t)$ and the QF peaks of $\tilde K_\downarrow(E\sim \Delta E_{\rm QF})$ originate from  the coherent dynamics of the  MEPs.
In order to deeply understand the microscopic origin of the QF, we try to find the  most relevant MEPs  that comprise the characteristic of the QF.
In the FIG. 2 in the main text, we consider the case when $N=30$ and $\gamma=10$.
%
%
Further analysis shows that the MEP with quantum number $\{I^{\rm e}_{\rm h},J^{\rm m}\}=\{-1,1\}$ is the most relevant one.
In  FIG. \ref{fs:K-E}, we plot  $\tilde K^{\rm MEPs}_\downarrow(\Delta E_{\rm  QF})$ with the characteristic energy difference $\Delta E_{\rm QF}$ between  the pair states $\Delta E_{\rm QF}$
In  the thermodynamic limit, we observe that $k^{\rm m}_{\rm p}=k^{\rm e}_{\rm p}$ because of  $I^{\rm m}_{\rm p}=I^{\rm e}_{\rm p}$. Thus  Eqs. (\ref{es:excitation-energy}) gives
\begin{eqnarray}
 \Delta E_{\rm QF} = | \varepsilon_{\rm c}(k^{\rm e}_{\rm h})- \varepsilon_{\rm s}(\lambda^{\rm e})+ \varepsilon_{\rm s}(\lambda^{\rm m})|.
\end{eqnarray}
In the thermodynamic limit, we further find from the TBA equations that $k^{\rm e}_{\rm h}=0$, $\lambda^{\rm m}=0$ and $\lambda^{\rm e}=-\infty$. Consequently the oscillation frequency (energy) of the QF is given by
\begin{eqnarray}
 \Delta E_{\rm QF}
 &=|\varepsilon_{\rm c}(0)|-|\varepsilon_{\rm s}(0)|.
\end{eqnarray}
Here  $\varepsilon_{\rm c}(0)<0$, $\varepsilon_{\rm s}(0)>0$ and $\varepsilon_{\rm s}(\pm\infty)=0$.
It follows that the result of Eq. (6) periodicity  of QF  in the main text
\begin{eqnarray}
 \tau_{\rm QF} = \frac{2\pi}{|\varepsilon_{\rm c}(0)|-|\varepsilon_{\rm s}(0)|}.
\end{eqnarray}
For the strong coupling limit  we have $|\varepsilon_{\rm c}(0)|-|\varepsilon_{\rm s}(0)| = E_{\rm F}[1-20\gamma/3+{\cal O}(\gamma^{-2})]$ that  gives
\begin{eqnarray}
 \tau_{\rm QF} = 2\pi t_{\rm F} \Big[1+ \frac{20}{3\gamma}+{\cal O}(\gamma^{-2})\Big].
\end{eqnarray}
These results were confirmed in the FIG.3 in the main text.

\begin{figure}[t]
\begin{center}
 \includegraphics[width=\linewidth]{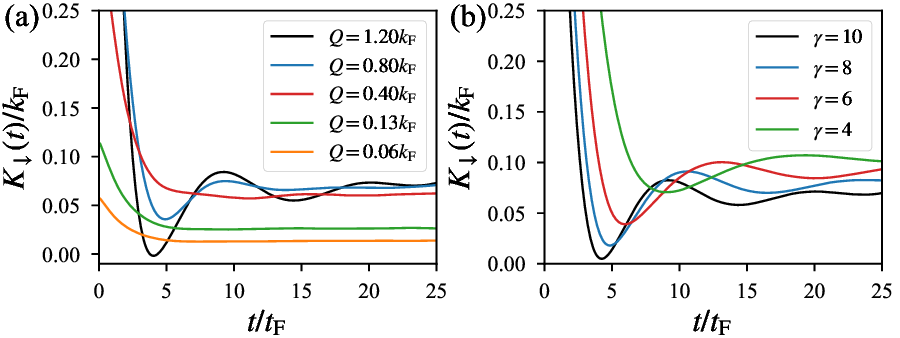}
\end{center}
\caption{QF of different conditions when $N=30$.
(a), $K_\downarrow(t)$ when $\gamma=10$ for different injected momentum $Q$.
(b), QF for different interaction strength $\gamma$ when $N=30$ when $Q=1.07k_{\rm F}$.
}
\label{fs:QF-different-case}
\end{figure}
%
%
%
%
%
%
In FIG. \ref{fs:QF-different-case} (a), we further demonstrate  the dynamics of impurity momentum for different initial momenta, ranging from  $Q< k_{\rm F}$   to $Q> k_{\rm F}$.
 It is showed that the saturated momentum approximately approaches to  the same value, but the oscillation amplitude increases when the $Q$ becomes larger.
When  $Q$ is small,  the QF no longer appears and the saturated momentum gradually turns to zero as  decreasing the  $Q$.
 In view of the fast decay process of $K_\downarrow(t)$, we observe that the momentum of the impurity decays faster when $Q$ becomes lager.
When $Q$ is large, $K_\downarrow$ even reach a negative value after the faster decay.
When the impurity is injected into the medium, the density of the medium in front of the faster moving impurity increases quickly so that quantum  friction between  the impurity and medium increases  quickly.
When the initial momentum $Q$ is larger than a critical value, the density of the medium in front of the impurity can be so dense such that  the impurity rebounds back from it.
 In FIG. \ref{fs:QF-different-case} (b), we demonstrate the interaction effect in the faster decay process and  the oscillation period.
From the  QF periodicity Eq. (6) in the main text, we observe  that the periodicity $\tau_{{\rm QF}}$ increases wen the interaction  $\gamma$ decreases,  see FIG. 3 in the main text.
%

\section{S5. Quantum revival}

\begin{figure}[t]
	\begin{center}
		\includegraphics[width=1\linewidth]{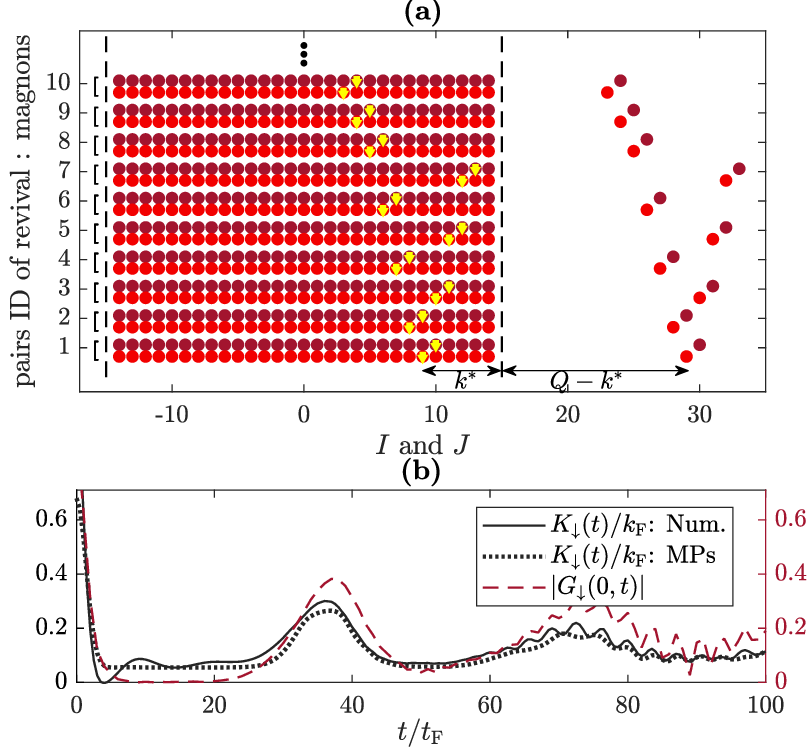}
	\end{center}
	\caption{
			The large weight pairs for quantum revival dynamics. (a) The red dots stand for the  quantum numbers of charges  $\boldsymbol{I}$. Whereas  the yellow arrows $\downarrow$ indicates  the quantum number $J$. The orders of these projected states are  the magnon states with  large  values of sum rule weights $|A_{\alpha}|^2+|A_{\alpha'}|^2$  used in Eq. (4) in the main text.
			The positions of the down spins in the pair are adjacent. Here  we present  the top 10 of such pairs of magnon-like projected states.
			We define the distance of the quantum number  $J$ to the Fermi surface as $k^*$.
			(b) The quantum revival of $K_{\downarrow}(t)$ is  obtained from the numerical result (black solid line) and from the propagator (red dashed line) for a plane wave impurity, showing   a good agreement with the result (the black dotted line) obtained from the states of MPs in (a). Here the particle number $N = 30$, interaction strength $\gamma = 10$ and the initial momentum $Q = 1.33k_{\rm F}$.}
	\label{fs:mps}
\end{figure}

Now we proceed to discover a microscopic origin of  the quantum revival from both 
$K_\downarrow(t)$
and $\tilde K_\downarrow(E)$.
The first peak of $\tilde K_\downarrow(E)$ in FIG. \ref{fs:K-E} shows that  the frequency is the  energy difference between the states in a  magnon pair  with nearest sequency quantum number $I^{\rm m}_{\rm p}$, see our discussion  in the beginning of Sec. S3.
Similar to the analysis on the QF,  here we further show that the first magnon pair illustrated in  FIG.\ref{fs:magnon-exciton} (a) determines  the position of the first peak of the  $\tilde K_\downarrow(E)$.
This is the most prominent pair of the magnon-like  states for the dynamics of the quantum revival.
Such a  pair of the magnon-like states  show the largest  weight of $|A_\alpha|^2$, see FIG. \ref{fs:magnon-exciton} (a) and  FIG. \ref{fs:mps}.

\begin{figure}[t]
\begin{center}
 \includegraphics[width=1\linewidth]{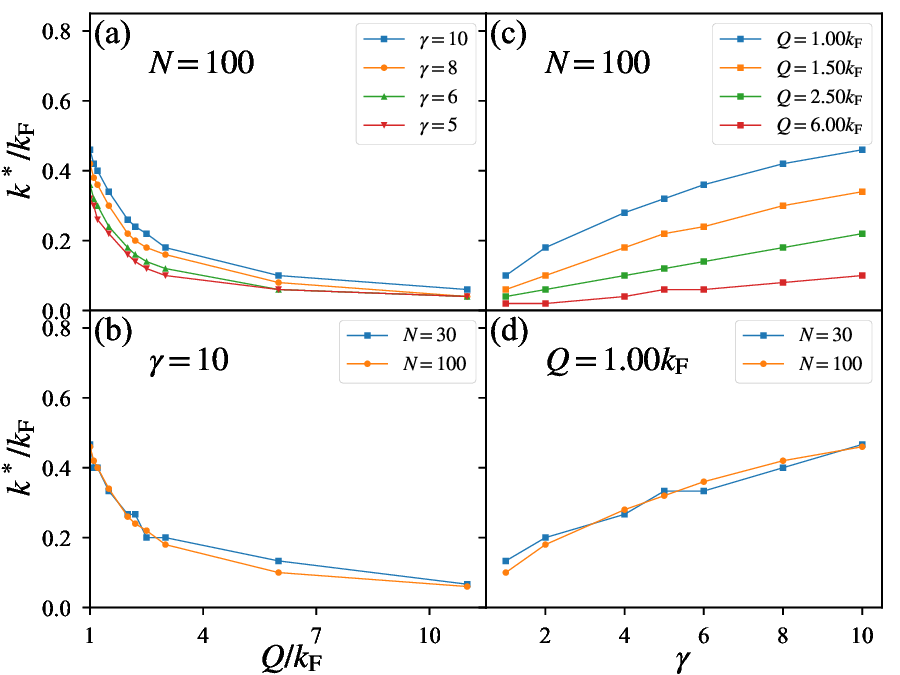}
\end{center}
 \caption{Numerical determination  of the parameter  $k^*$. (a) The parameter $k^*$ v.s. the initial impurity momentum shows the dependence of the interaction.
 (b) The parameter $k^*$ v.s. the initial impurity momentum shows the independence of the total particle number.  (c) The parameter $k^*$ v.s. the interaction strength shows the dependence of the initial momentum $Q$. (d) The parameter $k^*$ v.s. the interaction strength  shows  the independence of the total partial number.}
\label{fs:k-star}
\end{figure}

The magnon-like states are denoted by  $\{I^{\rm m}_{\rm p}, J^{\rm m}\}$ in Eqs. (\ref{es:mag-exc-quan-num}).
We denote the quantum numbers of the two states in the most prominent pair as $\{I_1, J_1\}$ and $\{I_2, J_2\}$, respectively,  leading to the  largest weight $|A_\alpha|^2$.  More precisely,
\begin{eqnarray}
 I_2 = I_1 \pm \Delta I,~~~J_2 = J_1 \pm \Delta J,
\end{eqnarray}
following which we have
\begin{eqnarray}
 k_2 = k_1 \pm \Delta k,~~~\lambda_2 = \lambda_1 \pm \Delta \lambda,
\end{eqnarray}
namely  $\Delta I=\Delta J=1$,  where the $k_{1,2}$ ($\lambda_{1,2}$) here are the corresponding wave numbers (rapidities) of $I_{1,2}$ ($J_{1,2}$).
The energy difference of the two states in this prominent pair gives the quantum revival $\Delta E_L$.
From Eq. (\ref{es:excitation-energy}), we have
\begin{eqnarray}
 \Delta E_L
 && =  \lim_{L\to\infty} |\varepsilon_{\rm c}(k_2) - \varepsilon_{\rm c}(k_1)
 + \varepsilon_{\rm s}(\lambda_{2}) - \varepsilon_{\rm s}(\lambda_1)|
 \nonumber\\
 &&= \lim_{L\to\infty}   | \Delta k \varepsilon_{\rm c}'(k_1)
  + \Delta \lambda \varepsilon'_{\rm s}(\lambda_1)|
 \nonumber\\
 &&= \lim_{L\to\infty}   \Big|
    \frac{\Delta k}{\Delta I} \Delta I\varepsilon_{\rm c}'(k_1)
  + \frac{\Delta \lambda}{\Delta J}\Delta J \varepsilon'_{\rm s}(\lambda_1)\Big|.
\end{eqnarray}
Moreover, we define
\begin{eqnarray}
 \begin{split}
  &\lim_{L\to\infty} \frac{\Delta I}{L \Delta k} = \rho_{\rm c}(k) + \rho^{\rm h}_{\rm c}(k),\\
  &\lim_{L\to\infty} \frac{\Delta J}{L \Delta \lambda} \Delta J = \rho_{\rm s}(k) + \rho^{\rm h}_{\rm s}(k).
 \end{split}
\end{eqnarray}
Then the characteristic  energy of quantum revival can be given by
\begin{equation}
 \Delta E_L
 =  \Delta p |  v_{\rm p}(Q-k^*) -  v_{\rm s}(k^*)|,
\end{equation}
with $k^*= k_{\rm F}-\frac{2\pi}{L} J_1$. Here $\Delta p = 2\pi/L$ and sound  velocities
\begin{eqnarray}
 &&
 v_{\rm p}\left(p\right)|_{p=(\frac{2\pi}{L}I_1-k_{\rm F})}=\frac{\varepsilon_{\rm c}'(k_1)}{2\pi [\rho_{\rm c}(k_{1})+\rho^{\rm h}_{\rm c}(k_1)]},
 \nonumber\\
 &&
 v_{\rm s}\left(p\right)|_{p=(k_{\rm F}-\frac{2\pi}{L}J_{1})}=\frac{\varepsilon_{\rm s}'(\lambda_1)}{2\pi [\rho_{\rm c}(\lambda_1)+\rho^{\rm h}_{\rm s}(\lambda_1)]}.
 \nonumber
\end{eqnarray}
Consequently, we find
\begin{eqnarray}
 \Delta E_{\rm L}
 && =  \frac{2\pi}{L} |  v_{\rm p}(Q-k^*) -  v_{\rm s}(k^*)|
 \nonumber\\
 && =  \frac{2\pi}{L} (  v_{\rm p}(Q-k^*) -  v_{\rm s}(k^*)),
\end{eqnarray}
where  $v_{\rm p}(p)$ is always larger than $v_{\rm s}(p)$.
This remarkably gives the period of quantum revival  Eq. (7) in the main text, namely $\tau_L = 2\pi /\Delta E_L$,
\begin{equation}
 \tau_{L} = \frac{N}{[v_{\rm c}(Q-k^*)-v_{\rm s}(k^*)]n} = \frac{L}{v_{\rm c}(Q-k^*)-v_{\rm s}(k^*)}.
\end{equation}
In this paper, $k^*$  was  calculated numerically based on the BA equations.
We also observe that $k^*$ is subject to  the impurity initial momentum $Q$ and interaction strength  $\gamma$, see   FIG. \ref{fs:k-star}.
However, $k^*$ dose not change  obviously   with respect to  $N$.
The revival dynamics of the supersonic impurity reveals the reflection of the collective excitations with respect to the finite-size effect.

\section{S6.  Supersonic impurity with a Gaussian  wave packet}

\begin{figure}[t]
\begin{center}
 \includegraphics[width=\linewidth]{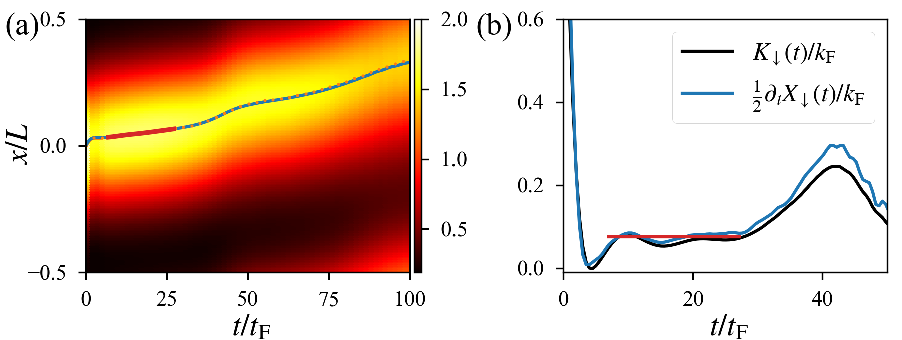}
\end{center}
\caption{QF and quantum revival in the injected Gaussian wave packet  impurity.
(a), The snaking signature occurs in the evolution of the  density distribution  $\braket{\hat{\Psi}^{\dagger}_{\downarrow}(x,t)\hat{\Psi}^{}_{\downarrow}(x,t)}$.
The blue line shows the motion of the mass center, $X_\downarrow(t)$, while the orange dots are the motion of wave packet center (position with maximum density). 
(b) The  evolution of  impurity momentum (black line) coincides with the evolution of the  wave packet center which is defined by $K_\downarrow(t)=\frac 12\partial_t X_\downarrow(t)$.
In a time interval  $7t_{\rm F}<t<27t_{\rm F}$, the average momentum (speed) of the mass center is  showed by the red straight line, which  matches  the saturated momentum $k_{\rm s}$ of the QF flutter.
In this figure, we set  $\gamma=10$, $Q=1.2k_{\rm F}$ and $a_0=0.21L$.
}
\label{fs:wave-packet}
\end{figure}

%
%
%
%
%
%
Now we consider a more realistic supersonic impurity with a Gaussian wave packet injected into the medium of bosonic liquid.
The impurity wave packet is given by
\begin{eqnarray}
 \phi_{\downarrow}(x)={\rm e}^{{\rm i}Qx}{\rm e}^{-(x/a_0)^2/2},
\end{eqnarray}
where $a_0$ is the width of the wave packet.
%
%
With the help of this injected wave packet, we calculate the evolution of the density
distribution of spin-down impurity, namely $\braket{\Psi^\dag_\downarrow (x,t)\Psi^{}_\downarrow(x,t)}$ ,  giving the result showed  in FIG. \ref{fs:wave-packet} (a).
The blue line in FIG. \ref{fs:wave-packet} (a) is the mass center of the impurity, showing a novel feature of   quantum snaking behavior.
More interesting to see that  the  snaking periodicity is  the same as the quantum revival period $\tau_L$.
We also observe that the oscillation dynamics of the QF flutter  of the impurity momentum also solely  appears only for  $Q \gtrsim k_{\rm F}$, see the black line FIG. \ref{fs:wave-packet} (b).
The microscopic origin of the QF and quantum revival here are the same as what we have found in the case with a plane wave injected into the bosonic medium.
%

%
In fact, the impurity momentum can be measured from the motion of the  mass center of the impurity
\begin{eqnarray}
 X_\downarrow(t)
 &&
 = \frac{\bra{\varPhi_{\rm I}} \hat x(t) \ket{\varPhi_{\rm I}}}{\braket{\varPhi_{\rm I}}},
\end{eqnarray}
where $\hat x(t)= {\rm e}^{{\rm i}\hat Ht} \hat x {\rm e}^{-{\rm i}\hat Ht}$ and $\hat x$ is the coordinate operator of the impurity.
Then
\begin{eqnarray}
 \partial_t X_\downarrow(t)
 &&
 = \frac{\bra{\varPhi_{\rm I}} {\rm e}^{{\rm i}\hat Ht} [\hat H \hat x - \hat x \hat H]{\rm e}^{-{\rm i}Ht} \ket{\varPhi_{\rm I}}}{\braket{\varPhi_{\rm I}}}
 \nonumber \\
 &&
 = \frac{\bra{\varPhi_{\rm I}} {\rm e}^{{\rm i}\hat Ht} {\rm i}[\hat p^2 \hat x -  \hat x \hat p^2]{\rm e}^{-{\rm i}Ht} \ket{\varPhi_{\rm I}}}{\braket{\varPhi_{\rm I}}},
 \nonumber
\end{eqnarray}
where $\hat p$ is the momentum operator of the impurity.
As such, we have that $ \hat x \hat p^2 = \hat p^2x  +2{\rm i} \hbar \hat p$.
It follows
\begin{eqnarray}
 \partial_t X_\downarrow(t)
 &&
 = 2\hbar \frac{\bra{\varPhi_{\rm I}} \hat K_\downarrow(t)  \ket{\varPhi_{\rm I}}}{\braket{\varPhi_{\rm I}}}
 = 2\hbar K_\downarrow(t).
 \nonumber
\end{eqnarray}
As mentioned in the main text, we set $\hbar=1$ here, and then
\begin{eqnarray}
 \frac12\partial_t X_\downarrow(t) = K_\downarrow(t).
\end{eqnarray}

We plot the $\frac12\partial_t X_\downarrow(t)$ ( blue line) in FIG. \ref{fs:wave-packet} (b).
The motion of the mass center  of the impurity  $\frac12\partial_t X_\downarrow(t)$  surprisingly coincides with the evolution of the impurity momentum $K_\downarrow(t)$.
A slight discrepancy between them is mainly because of  the finite size effect.
 In addition,  the sum rules in numerical calculations of $K_\downarrow(t)$ was not token up to $100\%$.
Therefore,  in a realistic experiment,  the QF and quantum revival behaviors can be observed by using the motion of the mass center $X_\downarrow(t)$.
Where we also  plot the saturated momentum of QF as the red line in FIG. (\ref{fs:wave-packet}) (b).

\end{document}